\documentclass[%
reprint,
amsmath,amssymb,
aps,
pra,
]{revtex4-2}

\usepackage[utf8]{inputenc}
\usepackage{color}
\usepackage{amsmath}
\usepackage{graphicx}
\usepackage{hyperref}
\usepackage{dcolumn}
\usepackage{bm}
\usepackage{wrapfig}
\usepackage{subfigure}
\usepackage{ucs}
\usepackage{lineno}
\usepackage{epsfig}
\usepackage{psfrag}
\usepackage{epstopdf}
\usepackage{float}
\usepackage{cuted}
\usepackage{flushend}
\usepackage{parcolumns}
\usepackage{lipsum,appendix}

\makeatletter
\@ifundefined{textcolor}{}
{
	\definecolor{BLACK}{gray}{0}
	\definecolor{WHITE}{gray}{1}
	\definecolor{RED}{rgb}{1,0,0}
	\definecolor{GREEN}{rgb}{0,1,0}
	\definecolor{BLUE}{rgb}{0,0,1}
	\definecolor{CYAN}{cmyk}{1,0,0,0}
	\definecolor{MAGENTA}{cmyk}{0,1,0,0}
	\definecolor{YELLOW}{cmyk}{0,0,1,0}
}

\begin{document}

\preprint{APS/123-QED}

\title{Robust quantum entanglement and teleportation in the  tetrapartite spin-1/2 square clusters: Theoretical study on the effect of a cyclic four-spin exchange}

  \author {Hamid Arian Zad$^{1,2}$, Azam Zoshki$^{2}$,  Nerses Ananikian$^{1,3}$, Michal Ja\v{s}\v{c}ur$^{2}$}
\affiliation{$^{1}$A.I. Alikhanyan National Science Laboratory, 0036, Yerevan, Armenia}
\affiliation{$^{2}$Department of Theoretical Physics and Astrophysics, Faculty of Science, P. J. S\v{a}f{\' a}rik University, Park Angelinum 9, 041 54 Ko\v{s}ice, Slovak Republic}
 \affiliation{ $^{3}$CANDLE Synchrotron Research Institute, Acharyan 31, 0040 Yerevan, Armenia}%



\date{\today}

\begin{abstract}
The whole entanglement measure so-called geometric $\Pi_4$ average of tangles and bipartite entanglement of the antiferromagnetic
 spin-1/2 XXX Heisenberg model on a tetranuclear square cluster with cyclic four-spin interaction are rigorously examined by the help of thermal negativities. The model comprises two nearest-neighbor exchange couplings $J_1$ and $J_2$ such that $J_1\gg J_2$. When the cyclic four-spin exchange is zero, the maximum value of whole entanglement $\Pi_4$ is achieved at low enough temperatures and relatively high magnetic fields $(B\approx J_1)$.
Also, maximum bipartite entanglement between pair spins with exchange coupling $J_1$ is achievable at high temperature and high magnetic field. This quantity remains alive for sufficiently high temperature and high magnetic field values comparable with the relevant exchange coupling $J_1$. 
A nonzero value of the cyclic four-spin exchange notably enhances the degree of the whole entanglement, while it weakens the bipartite entanglement degree. We demonstrate that the whole entanglement reaches an unconventional minimum at a special parameter region of cyclic four-spin exchange almost ten order of magnitude smaller than $J_1$, where the system is in a quantum antiferromagnetic state.
 The real complex $[\text{Cu}_4\text{L}_4(\text{H}_2\text{O})_4](\text{ClO}_4)_4$ as  a strong antiferromagnetic tetranuclear square compound provides us an experimental representative
to estimate the strength of the whole and bipartite entanglements at high enough temperature.
 It is demonstrated that the entanglement negativities of this complex are yet depend on the considered cyclic four-spin interaction even though its magnitude is significantly smaller than $J_1$.
We realize that the possibility of quantum teleportation through a couple of discrete $\text{Cu}_4^{\text{II}}$ complexes would solely happen for the magnetic fields lower than a critical value. 
 It is also demonstrated  that enhancing the cyclic four-spin exchange results in diminishing the average fidelity. 
Moreover, we conclude that the quantum Fisher information investigated for the pair of spins with exchange interaction $J_1$ reaches its minimum close to the critical magnetic field at which a ground-state phase transition occurs from highly entangled state to separable one.

\end{abstract}

\maketitle

\section{Introduction}
 
Low-dimensional spin systems are of particular interest as magnetic materials for recent developments in 
 quantum information technologies and molecular spintronics \cite{Kahn1993,Benelli2015,Gaita2019,Yurtaeva2019,Kintzel2018,Camarero2009}.
 Scientists have recently been interested in the application of antiferromagnets by which the information could be encoded in the direction of the N{\'e}el vector. This is due to the fact that antiferromagnetic materials are not only able to write information much faster but also are very stable against external fields \cite{Bodnar2018}.
 
The characterized tetranuclear copper(Cu$^\text{II}$) complexes as exchange-coupled systems have attracted a great deal
of attention since past decade owing to their potential applications in a number of branches such as 
coordination polymers \cite{Sutradhar2020,Tandon2020}, molecular magnetism \cite{Kahn1993}, high-temperature superconductors \cite{Burkhardt2008}, string theory \cite{Borsten2010} and quantum information processing \cite{Bodnar2018}. 
Among them, strong antiferromagnetically tetranuclear Cu$^\text{II}$ complexes  have particularly studied due to bear magneto-structural properties and reveal zero-temperature phase transitions between possible ground states. 
For example, magnetic properties and crystal structure of a tetranuclear cubane-type Cu$^\text{II}$ complex based on a 
hexa-coordinating $\text{N}_2\text{O}_4$ ligand have been reported in Ref. \cite{Maity2010}.
Next, S. C. Manna {\it et al.}  synthesized and characterized two tetranuclear closed-cubane like complexes $[\text{Cu}_4(\text{L}^1)_4]\cdot 3(\text{H}_2\text{O})$ and $[\text{Cu}_4(\text{H}_2\text{L}^2)_4(\text{H}_2\text{O})_4]$ \cite{Manna2017}, and based on the Tercero {\it et al.} \cite{Tercero2006} theoretical tasks, they determined the nature and magnitude of exchange couplings between Cu atoms of these complexes that are correlated with the $\text{Cu}-\text{O}-\text{Cu}$ angle. 

The structural and magnetic properties of the ferromagnetically coupled tetranuclear copper $\text{Cu}_4^{\text{II}}$ square complex $[{\text{Cu}(\text{H}_2\text{L})}_4(\text{THF})]$  were experimentally investigated in Ref. \cite{Salmon2014}. The spin arrangement of this special complex in the crystal lattice leads to the formation of square structure. 
From the expected values for the six coupling constants, four intracubane (square sides) interactions and two intercubane (diagonals) interactions,  intercubane interactions are negligible and one can remove them from the
magnetic coupling pathway scheme. Owing to this fact, intracubane interactions have been often observed as strong interactions
 in the fitting procedure, therefore the main contribution being the coupling along the square sides, and thus the diagonal contribution should be taken in to account with caution.

The crystal structure and magnetic properties of a tetranuclear Cu$^\text{II}$ complex containing the 2-(pyridine-2-yliminomethyl)-phenol ligand have been studied in detail by S. Giri {\it et al.} in Ref. \cite{Giri2011}. Indeed, they synthesized a pure
 $\mathrm{Cu}^\mathrm{II}$ compound, $[\mathrm{Cu}_4\mathrm{L}_4(\mathrm{H}_2\mathrm{O})_4](\mathrm{ClO}_4)_4$, that is a  Schiff base ligand with the perchlorate counter ion.
 The structure of this complex includes discrete planar tetranuclear $[\mathrm{Cu}_4\mathrm{L}_4(\mathrm{H}_2\mathrm{O})_4]^{4+}$ cations together with perchlorate anions. 
The four copper atoms that constitute a unit tetrameric block are crystallographically equivalent and are five-coordinate with square pyramidal environments. 
Fitting process of the magnetic susceptibility data with the experiment indicated that  the compound has a strong antiferromagnetic, $J_1/k_B=-638\; \text{K}$, exchange interaction between diphenoxo-bridged
 Cu$^\text{II}$ centers, as well as, a moderate antiferromagnetic, $J_2/k_B=-34\; \text{K}$, interaction between 
 $\text{N}-\text{C}-\text{N}$ bridged Cu$^\text{II}$ centers. 
Such planar cyclic $\mathrm{Cu}_4\mathrm{O}_4$ core structures involve a $\mathrm{Cu}_2\mathrm{O}_2$ bridging moiety
 and have attracted a considerable interest due to be momentous in seeking molecular precursor compounds for synthesizing 
 high-temperature superconductors \cite{Trepanier1992,Fielden2006,Plass2008}.

  Non-classical correlations between different parties of a quantum system can be described by quantum entanglement where it plays as an important hallmark that is able to separate quantum systems from their classical counterparts. 
To quantify the entanglement degree of a given bipartite quantum state two popular measures: Concurrence and
negativity can be applied \cite{Wootters2000,Plenio2005,Vidal2002}.
 The negativity \cite{Vidal2002} as a measure of either bipartite or multipartite entanglement can be considered according to the necessary condition Peres--Horodecki separability criterion \cite{Peres1996,Horodecki1996}. This criterion is called the PPT (positive partial transpose) criterion and determines the separability of a two quantum mechanical subsystems A and B of a global system (AB). 
Owing to this fact, the quantity negativity has been widely employed to measure the degree of the thermal entanglement of the
multipartite Heisenberg spin models \cite{Guo2007,Guo2008,Xu2014,Wang2008,Han2017}.  
  
 Motivated by the complex, $[\text{Cu}_4\text{L}_4(\text{H}_2\text{O})_4](\text{ClO}_4)_4$ \cite{Giri2011} and its likely application in superconducting quantum computers at high temperatures, in the present work, we consider this compound as an experimental sample and  proceed ourselves to theoretically investigate various quantum protocols such as bipartite and tetrapartite entanglement negativities \cite{Wootters1998,Nielsen2000,Vedral2002,Horodecki2009},  quantum teleportation \cite{Bennett1993,Huang2011,Arian2021PhysE,Rojas2019,Fouokeng2020}, and quantum Fisher information (QFI) \cite{Bakmou2019,Slaoui2019,Jafari2020,Marzolino2017,Liu2020,Ye2020,Ye2018} over the room temperature scales.
The QFI is a useful tool that makes the possibility of estimating the quantum metrology of a spin system and can be plausibly connected to the quantum phase transition at higher temperatures. Hence, we disclose anomalous behavior of these quantities nearby the critical points at which a ground-state phase transition occurs. 
 
The most important feature of spin systems is the type of interaction between their ingredients. It has already been pronounced that in addition to the bilinear exchange interaction, cyclic four-spin interactions known as the ring exchange terms are momentous in the subject \cite{Honda1993,Brehmer1999,Senthil2001,Reischl2002,Balents2002,Capponi2013,Eggert2017,Arian2017,Michal2019,Arian2021}.
In Ref. \cite{Calzado2003} the cyclic four-spin exchange term was detected in some cuprates and calculated theoretically.
The authors of Ref. \cite{Buhler2011} investigated the susceptibility and specific heat of a two-leg $S-1/2$ ladder possessing cyclic four-spin exchange and realized that a small cyclic exchange interaction remarkably affects the thermodynamics of the model even at high temperature.
 However, realizing the effects of ring exchange interaction on the magnetic and quantum properties of small cluster magnets specially tetranuclear Cu$^\text{II}$ complexes is yet an open question and has not been seriously attempted to answer.
 Therefore, in the present work we theoretically address the question: How the bipartite and tetrapartite quantum entanglements of the spin-1/2 antiferromagnetic Heisenberg model on a tetranuclear 
  square compound (i.g., $[\text{Cu}_4\text{L}_4(\text{H}_2\text{O})_4](\text{ClO}_4)_4$) are changed by cyclic four-spin exchange interaction?

This paper is organized as follows. In Sec. \ref{model} we describe the spin-1/2 antiferromagnetic Heisenberg XXX model on a tetranuclear cluster magnet with the cyclic four-spin interaction in the presence of an external magnetic field. Next, we classify its eigenenergies with the corresponding eigenstates and the most interesting results for the ground-state phase diagram is figured, as well.
In Sec. \ref{results}, firstly we theoretically describe the influences of a cyclic four-spin interaction on the whole entanglement negativity of the model under consideration. Secondly, we consider the real complex  $[\text{Cu}_4\text{L}_4(\text{H}_2\text{O})_4](\text{ClO}_4)_4$ as an experimental example of the strong antiferromagnetic Heisenberg model on a tetranuclear square compound that has a very strong exchange coupling $J_1/k_B = -638\;\mathrm{K}$ and a moderate exchange coupling $J_2/k_B = -34\;\mathrm{K}$,  then we discuss the effects of a typical cyclic four-spin exchange on the whole entanglement and bipartite quantum entanglement between Cu ions with stronger antiferromagnetic interaction.
 In the final part of this section, by quantifying average fidelity, we  rigorously examine the possibility of the quantum teleportation through this model at high temperature. Furthermore, the behavior of QFI is compared with the average fidelity and the most stimulating findings are reported, as well.
 Finally, the concluding remarks are given in Sec. \ref{conclusions}.

\section{The model}\label{model}

First let us define the Hamiltonian of the spin-1/2 Heisenberg model on a tetranuclear square compound as
\begin{equation}
\mathcal{H} = \sum\limits_{i=1}^N{H}_i,
\end{equation}
where ${H}_i$ is the effective Hamiltonian of a tetrameric unit block, indicating a discrete four-spin system (see supporting information of Ref. \cite{Giri2011}).
\begin{figure}
\centering
\includegraphics[scale=0.6,trim=50 100 100 50, clip]{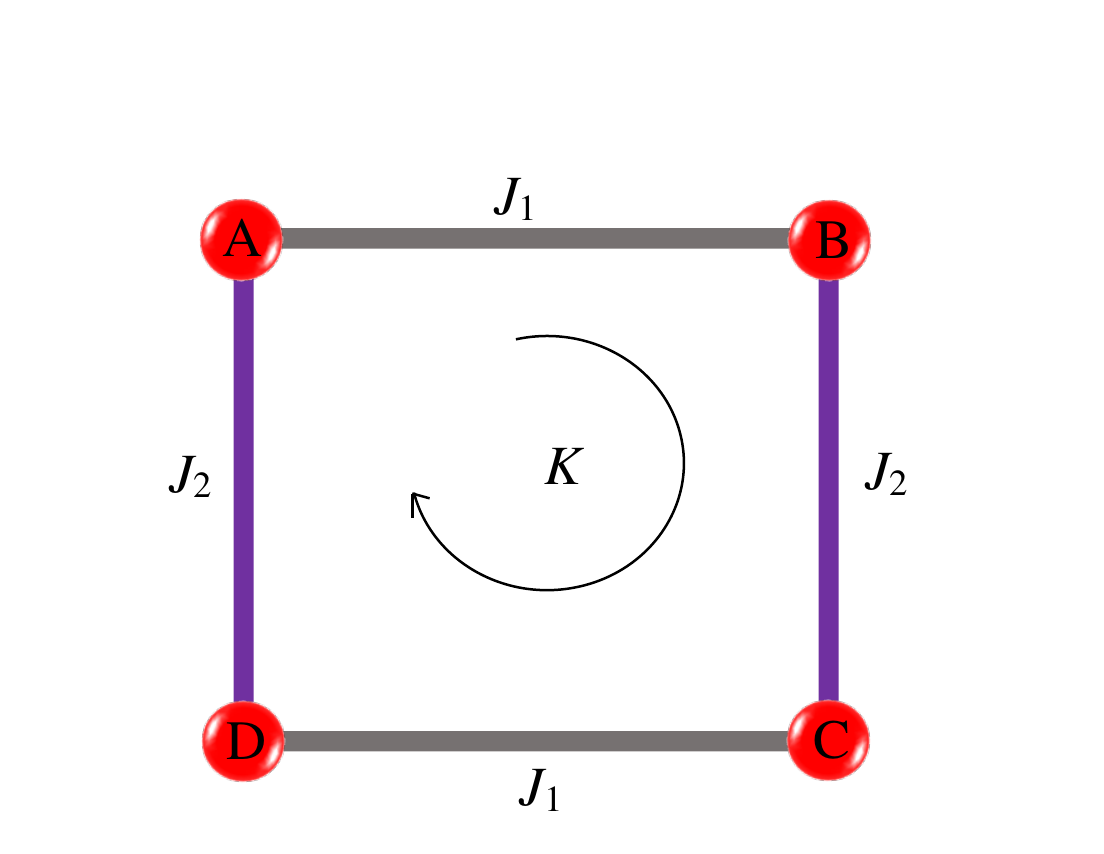}
\caption{Structural arrangement of the spin-1/2 Heisenberg model on a square compound with magnetic exchange interactions $J_1$, $J_2$ and cyclic four-spin term $K$.}
\label{fig:Model}
\end{figure}
We suppose that the form of Hamiltonian of all unit blocks are identical (see Fig. \ref{fig:Model}), therefore Hamiltonian $H={H}_i$ can be expressed as
\begin{equation}\label{Hamitonain}
\begin{array}{lcl} 
{H} =-J_{1}\left[{\boldsymbol S}_\mathrm{A}\cdot {\boldsymbol S}_\mathrm{B}
 +{\boldsymbol S}_\mathrm{C}\cdot {\boldsymbol S}_\mathrm{D}\right]
-J_{2}\left[{\boldsymbol S}_\mathrm{B}\cdot {\boldsymbol S}_\mathrm{C}+{\boldsymbol S}_\mathrm{A}\cdot {\boldsymbol S}_\mathrm{D} \right]\\
 \quad\quad\;+ K[({\boldsymbol S}_\mathrm{A}\cdot{\boldsymbol S}_\mathrm{B})({\boldsymbol S}_\mathrm{C}\cdot{\boldsymbol S}_\mathrm{D})+({\boldsymbol S}_\mathrm{A}\cdot{\boldsymbol S}_\mathrm{D})({\boldsymbol S}_\mathrm{B}\cdot{\boldsymbol S}_\mathrm{C})\\ 
\quad\quad\; - ({\boldsymbol S}_\mathrm{A}\cdot{\boldsymbol S}_\mathrm{C})({\boldsymbol S}_\mathrm{B}\cdot{\boldsymbol S}_\mathrm{D})]
  -\mu_B g B(S_\mathrm{A}^{z}+S_\mathrm{B}^{z}+S_\mathrm{C}^{z}+S_\mathrm{D}^{z}).
\end{array}
\end{equation}

where ${\boldsymbol S}_a\cdot {\boldsymbol S}_{b}=S_{a}^{x}S_{b}^{x}+S_{a}^{y}S_{b}^{y}+S_{a}^{z}S_{b}^{z}$
with $a,b=\{\mathrm{A,B,C,D}\}$. The coupling $J_i$ ($i=\{1,2\}$) describes the strength of the spin
interaction being  antiferromagnetic $J_i<0$ and ferromagnetic $J_i>0$. 
The local asymmetric formation of the four different chemical environments of the copper centers results in two different magnetic exchange interactions including $\{J_1, J_2\}$ correspond to the square sides that play the main role of exchange contribution \cite{Giri2011}.
$B$ is the external magnetic field which is considered in the $z$-direction. $K$ is the cyclic four-spin exchange interaction. 
$\mu_B$ is Bohr magneton, $g$ is Land{\' e} g-factor that motivated by the experimental representative considered in this work we set $g = 2.19$. $S_{i}^{\alpha}$ ($\alpha =x,y,z$) are the spin-1/2 operators with $\hbar =1$.

 By implementing the exact analytical diagonalization one can directly solve the model and derive the eigenvalues of the Hamiltonian
  $\mathcal{H}_i$ as
\linespread{0.1}
 \begin{widetext}
 {\small
\begin{equation*}
\begin{array}{lcl}
\left\{
  \begin{array}{@{}l@{}}
E_{1}=g\mu_BB-\dfrac{1}{2}J_1-\dfrac{1}{2}J_2+\dfrac{1}{16}K,\\[0.3cm]
E_{2}=g\mu_BB-\dfrac{1}{2}J_1+\dfrac{1}{2}J_2+\dfrac{1}{16}K,\\[0.3cm]
E_{3}=g\mu_BB+\dfrac{1}{2}J_1-\dfrac{1}{2}J_2+\dfrac{1}{16}K,\\[0.3cm]
E_{4}=g\mu_BB+\dfrac{1}{2}J_1+\dfrac{1}{2}J_2-\dfrac{7}{16}K,\\[0.3cm]
\end{array}\right. \quad\quad
\left\{
  \begin{array}{@{}l@{}}
E_{5}=\dfrac{1}{2}J_1+\dfrac{1}{2}J_2+\dfrac{5}{16}K-\dfrac{1}{2}\sqrt{\Lambda},\\[0.3cm]
E_{6}=\dfrac{1}{2}J_1+\dfrac{1}{2}J_2+\dfrac{5}{16}K+\dfrac{1}{2}\sqrt{\Lambda},\\[.3cm]
E_{7}=-\dfrac{1}{2}J_1-\dfrac{1}{2}J_2+\dfrac{1}{16}K,\\[0.3cm]
E_{8}=-\dfrac{1}{2}J_1+\dfrac{1}{2}J_2+\dfrac{1}{16}K,\\[0.3cm]
E_{9}=-g\mu_BB+\dfrac{1}{2}J_1-\dfrac{1}{2}J_2+\dfrac{1}{16}K,\\[0.3cm]
E_{10}=\dfrac{1}{2}J_1-\dfrac{1}{2}J_2+\dfrac{1}{16}K,
\end{array}\right. \quad 
\left\{
  \begin{array}{@{}l@{}}
E_{11}=\dfrac{1}{2}J_1+\dfrac{1}{2}J_2-\dfrac{7}{16}K,\\[0.25cm]
E_{12}=-g\mu_BB+\dfrac{1}{2}J_1+\dfrac{1}{2}J_2-\dfrac{7}{16}K,\\[.25cm]
E_{13}=-g\mu_BB-\dfrac{1}{2}J_1+\dfrac{1}{2}J_2+\dfrac{1}{16}K,\\[0.3cm]
E_{14}=-g\mu_BB+\dfrac{1}{2}J_1-\dfrac{1}{2}J_2+\dfrac{1}{16}K,\\[0.3cm]
\end{array}\right. \\[0.75cm]
\left\{
  \begin{array}{@{}l@{}}
E_{15}=-2g\mu_BB-\dfrac{1}{2}J_1-\dfrac{1}{2}J_2+\dfrac{1}{16}K,\quad\\[0.3cm]
E_{16}=2g\mu_BB-\dfrac{1}{2}J_1-\dfrac{1}{2}J_2+\dfrac{1}{16}K,\quad
\end{array}\right.

\end{array}
\end{equation*}%
}
\end{widetext}
by adopting $\Lambda=4J_1^2-4J_1J_2+4J_2^2+2J_1K+2J_2K+K^2.$
 The eigenvectors of the $\mathcal{H}_i$ in terms of standard basis are classified according to the total quantum spin number
 $S_\text{T}=S_\mathrm{A}+S_\mathrm{B}+S_\mathrm{C}+S_\mathrm{D}$  and its $z-$component $S^z_\text{T}$ as follows
 \begin{widetext}
\begin{equation*}
\begin{array}{lcl}
(S_\text{T}=1, S^z_\text{T}=-1):\left\{
  \begin{array}{@{}l@{}}
  \left\vert \psi_{1}\right\rangle =\left\vert 1000\right\rangle +\left\vert 0100\right\rangle +\left\vert 0010\right\rangle +\left\vert 0001\right\rangle,
\\[0.025cm]
\left\vert \psi _{2}\right\rangle =-\left\vert 1000\right\rangle-\left\vert 0100\right\rangle +\left\vert 0010\right\rangle +\left\vert
0001\right\rangle ,
\\[0.025cm]
\left\vert \psi _{3}\right\rangle =\left\vert 1000\right\rangle -\left\vert 0100\right\rangle -\left\vert 0010\right\rangle +\left\vert 0001\right\rangle,
\\[0.025cm]
\left\vert \psi _{4}\right\rangle =-\left\vert 1000\right\rangle+\left\vert 0100\right\rangle -\left\vert 0010\right\rangle+\left\vert 0001\right\rangle,
 \end{array}\right. \\[0.75cm]
\quad\quad\quad\quad(S_\text{T}=0):\;\left\{
\begin{array}{@{}l@{}}
\left\vert \psi _{5}\right\rangle =a(\left\vert 1100\right\rangle+\left\vert 0011\right\rangle)+
b(\left\vert 1010\right\rangle +\left\vert 0101\right\rangle)
-c(\left\vert 1001\right\rangle +\left\vert 0110\right\rangle),
\\
\left\vert \psi _{6}\right\rangle =d(\left\vert 1100\right\rangle+\left\vert 0011\right\rangle)-
e(\left\vert 1010\right\rangle +\left\vert 0101\right\rangle)
-f(\left\vert 1001\right\rangle +\left\vert 0110\right\rangle),
\\
\left\vert \psi _{7}\right\rangle=\left\vert 1100\right\rangle +\left\vert 1010\right\rangle+\left\vert 1001\right\rangle +\left\vert 0110\right\rangle+\left\vert 0101\right\rangle+\left\vert 0011\right\rangle ,
\\
\left\vert \psi _{8}\right\rangle =-\left\vert 1010\right\rangle+\left\vert 0101\right\rangle,
\\
\left\vert \psi _{9}\right\rangle=-\left\vert 1100\right\rangle +\left\vert 0011\right\rangle,
\\
\left\vert \psi _{10}\right\rangle =-\left\vert 1001\right\rangle+\left\vert 0110\right\rangle,
\end{array}\right. \\[0.75cm]

\;\;(S_\text{T}=1, S^z_\text{T}=1):\;\left\{
\begin{array}{@{}l@{}}
\left\vert \psi _{11}\right\rangle=\left\vert 1110\right\rangle +\left\vert 1101\right\rangle+\left\vert 1011\right\rangle+\left\vert 0111\right\rangle,
\\
\left\vert \psi _{12}\right\rangle =-\left\vert 1110\right\rangle +\left\vert 1101\right\rangle -\left\vert 1011\right\rangle+\left\vert 0111\right\rangle ,
\\
\left\vert \psi _{13}\right\rangle =-\left\vert 1110\right\rangle -\left\vert 1101\right\rangle+\left\vert 1011\right\rangle+\left\vert 0111\right\rangle ,
\\
\left\vert \psi _{14}\right\rangle =\left\vert 1110\right\rangle-\left\vert 1101\right\rangle-\left\vert 1011\right\rangle+\left\vert 0111\right\rangle ,
\\
\end{array}\right. \\[0.75cm]
(S_\text{T}=2, S^z_\text{T}=\pm2):\;\left\{
\begin{array}{@{}l@{}}
\left\vert \psi _{15}\right\rangle =\left\vert 1111\right\rangle,
\\
\left\vert \psi _{16}\right\rangle =\left\vert 0000\right\rangle,
\end{array}\right.
\end{array}
\end{equation*}
\end{widetext}
where 1 and 0 are nominated as spins-up and spins-down, respectively. $\left\vert 1\right\rangle$ and $\left\vert 0\right\rangle$ 
are the eigenbasis of the 2 by 2 Pauli matrix $\sigma^z=\text{diag}(1,-1)$.
Coefficients $\{a, b, c, d, e, f\}$ are so long polynomial functions of $B$, $J_1$, $J_2$ and $K$ that for simplicity we numerically obtain them in order to examine forthcoming quantities.

\subsection{Ground-state phase transition}\label{GPT}
The spin-1/2 tetranuclear square complex with Hamiltonian (\ref{Hamitonain}) exhibits four different ground states in the presence of an external magnetic field.
The ground-state phase diagram of the model involves two localized one spin-down phases, $\vert \psi_{11} \rangle$ and $\vert \psi_{13} \rangle$, the localized two spin-up phase $\vert\psi_5 \rangle$ and the fully polarized phase $\vert \psi_{15} \rangle$. These ground states can be represented by the form $\vert S_\text{T}, S_\mathrm{AB},S_\mathrm{CD}\rangle$ \cite{JozefAPPA2020,JozefM2020} as 
\begin{equation}\label{FourGS}
\begin{array}{lcl}
\vert\psi_5 \rangle=\vert {0,1,1} \rangle \quad \text{with}\quad M/M_\text{s}=0,\\[0.2cm]
\vert \psi_{11} \rangle=\vert {1,1,1} \rangle_\text{I}\quad \text{with}\quad M/M_\text{s}=\dfrac{1}{2},\\[0.2cm]
\vert\psi_{13} \rangle=\vert {1,1,1} \rangle_\text{II}\quad \text{with}\quad M/M_\text{s}=\dfrac{1}{2},\\[0.2cm]
\vert \psi_{15} \rangle=\vert 2,1,1 \rangle\quad \text{with}\quad M/M_\text{s}=1,
\end{array}
\end{equation}
where
$S_{\mathrm{AB}}=S_\mathrm{A}+S_\mathrm{B}$ and $S_{\mathrm{CD}}=S_\mathrm{C}+S_\mathrm{D}$ are spin numbers of two faced dimers.
$M=-(\partial{\mathcal{G}}/\partial{B})_{T}$ represents the total magnetization where $\mathcal{G}=-k_BT\ln Z$ is the Gibbs free energy with $Z$ being the partition function of the model. $T$ is the temperature, $k_B$ is Boltzmann constant and $M_\text{s}$ is the saturation magnetization.

\begin{figure*}[tbp]
\centering
\includegraphics[width=7.55cm, height=5.25cm]{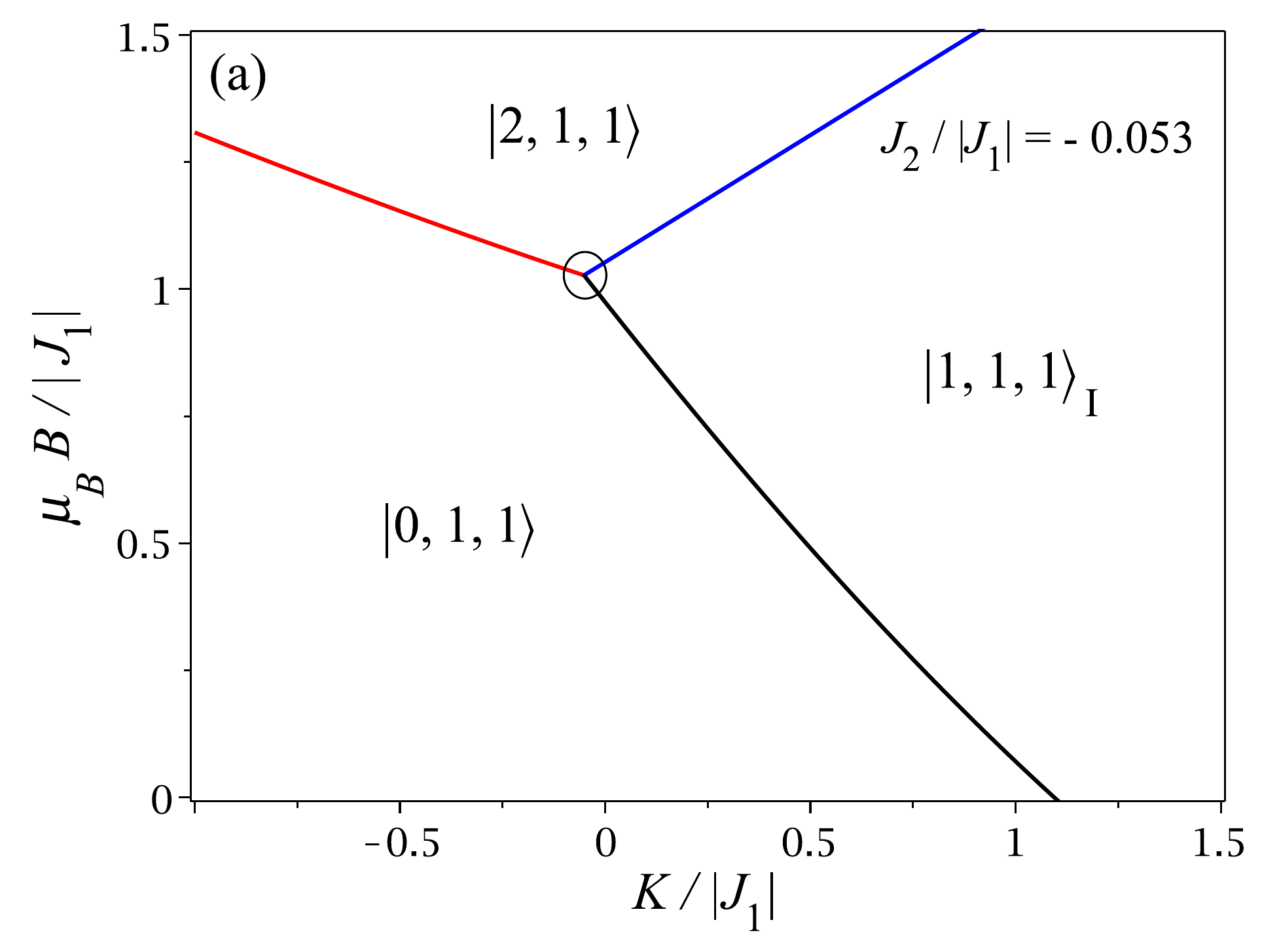}
\includegraphics[width=7.25cm, height=5.30cm]{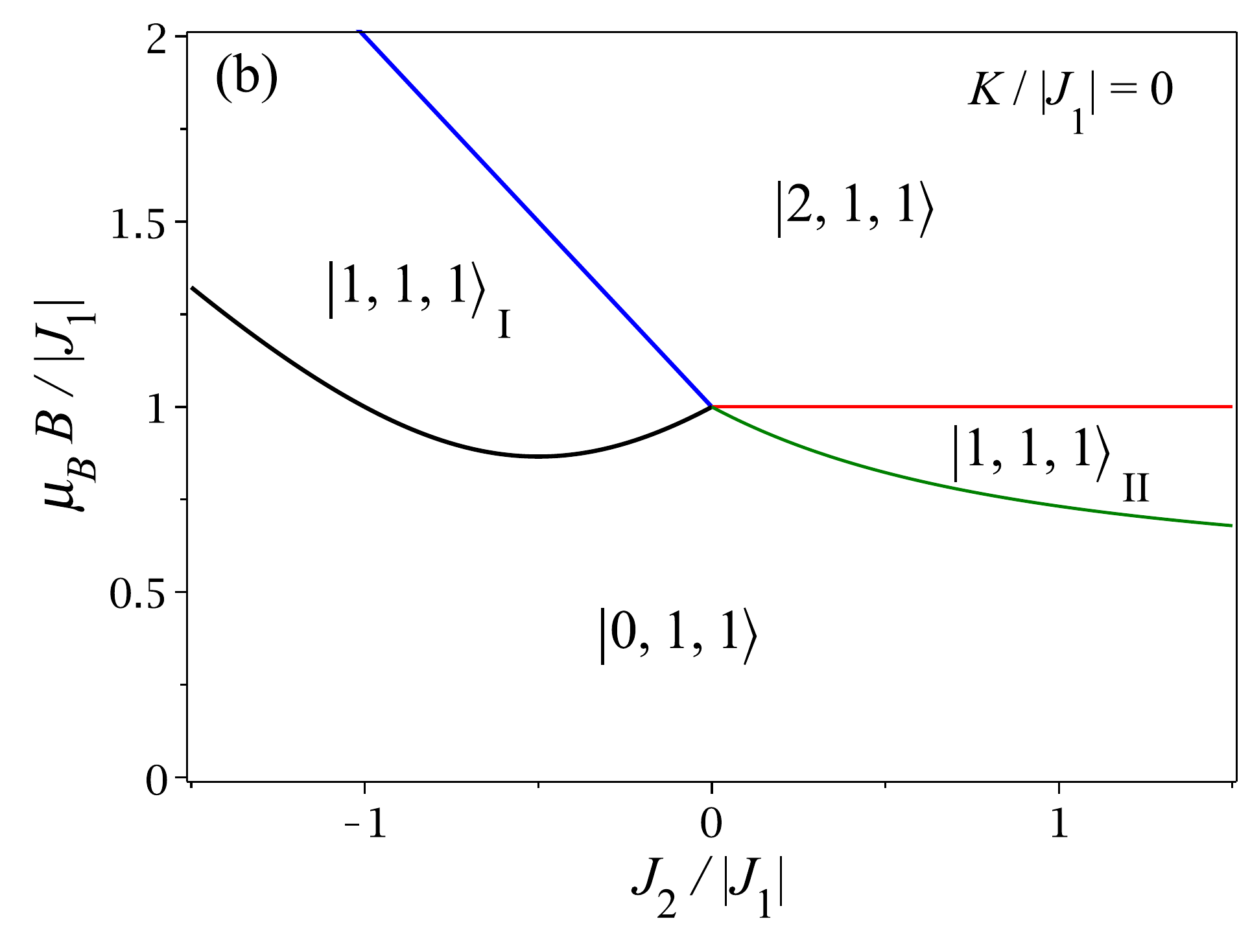}
\includegraphics[width=7.4cm, height=5.25cm]{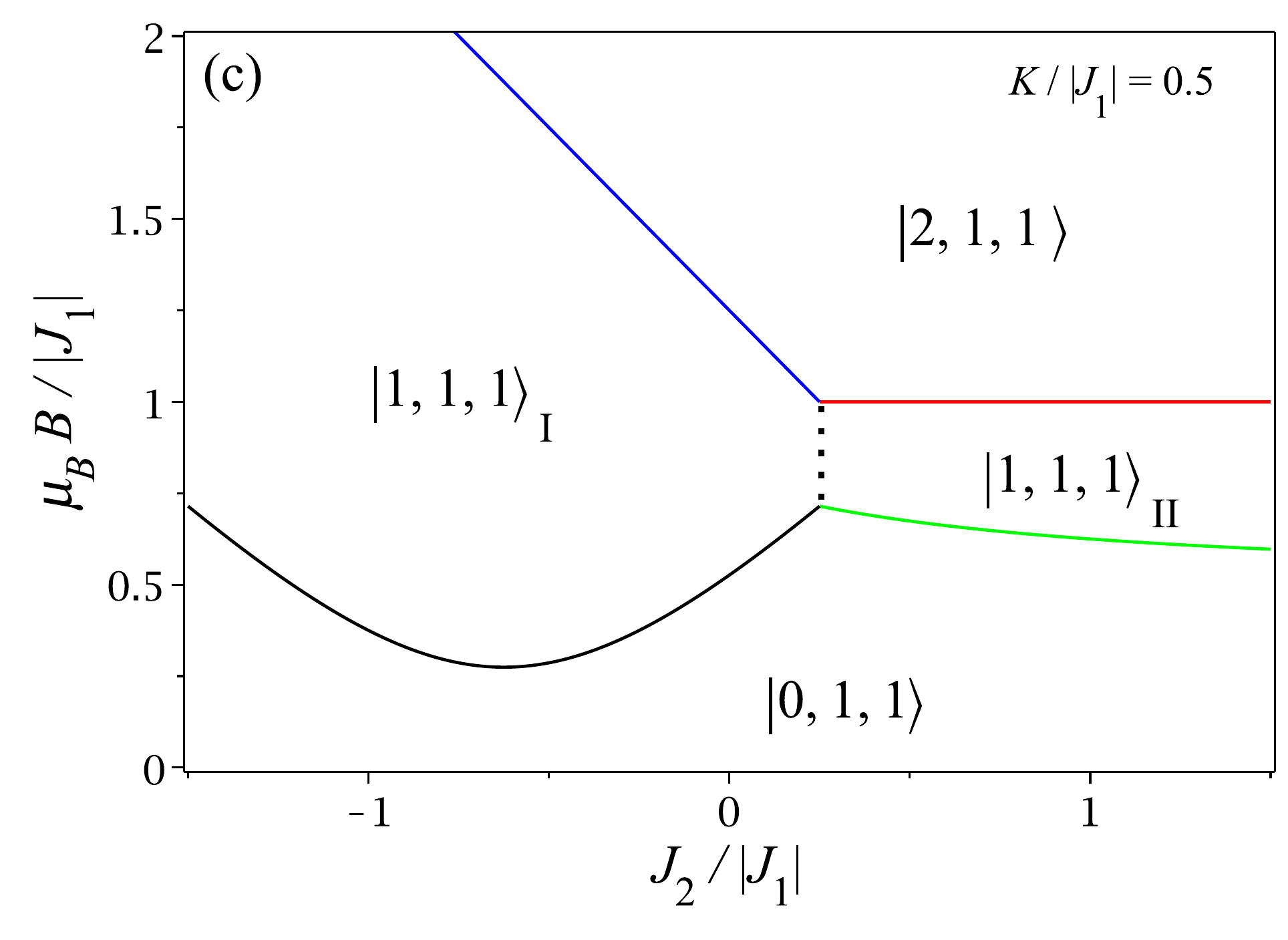}
\includegraphics[width=7.4cm, height=5.25cm]{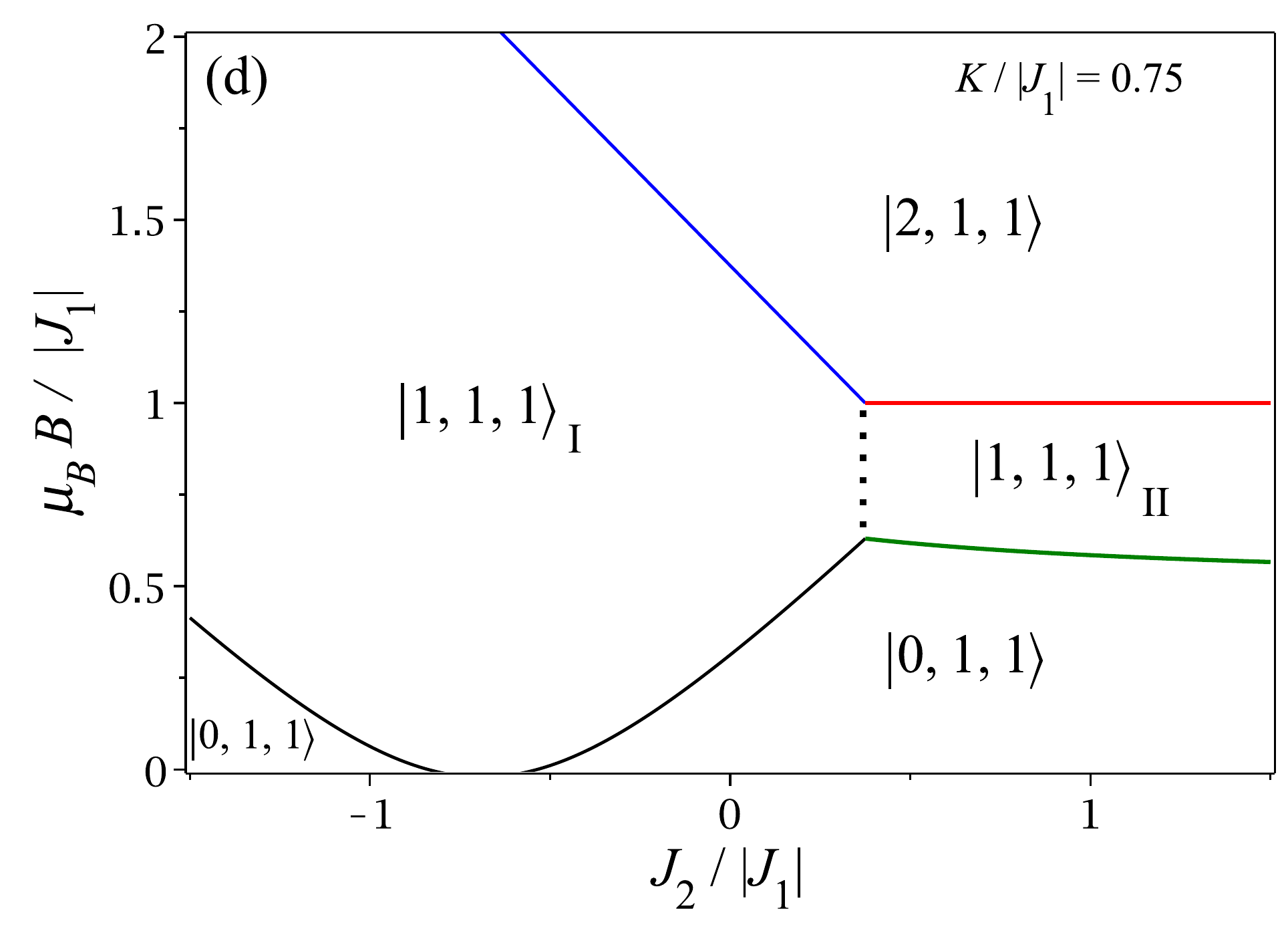}

\caption{(a) Ground-state phase diagram in the $\mu_BB/|J_1|-K/|J_1|$ plane by assuming fixed $J_2/|J_1| = - 0.053$ that indicates both exchange couplings are antiferromagnetic such that $|J_1|$ is assumed to be the energy unit and $J_2/|J_1|=-0.053$ being very weak antiferromagnetic with respect to the $|J_1|$. The triple point at which three different ground states coexist together is indicated by circle.
 Ground-state phase diagram in the $\mu_BB/|J_1|-J_2/|J_1|$ plane by assuming; (b) $K = 0$,
(c) $K/|J_1| = 0.5$, (d) $K/J|J_1| = 0.75$. Phase boundary between each two different states is depicted by solid line with different color. Vertical dotted line in lower panels shows the phase boundary between two ground states $\vert {1,1,1} \rangle_\text{I}$ and $\vert {1,1,1} \rangle_\text{II}$ with the same magnetization value $M/M_\text{s}=1/2$.}
\label{GSPT}
\end{figure*}

In Fig. \ref{GSPT} (a), we display the zero-temperature ground-state  phase diagram in the
$\mu_BB/|J_1|-K/|J_1|$ plane where weak antiferromagnetic coupling constant $J_2/|J_1| = -0.053$ is assumed.
We choose this value according to the experimental findings for the nature and strength of the exchange couplings reported in
 Ref. \cite{Giri2011}. 
It is noteworthy that here we evaluate parameters $\{\mu_BB,\; J_2,\; K\}$ with respect to the $|J_1|$ (where $J_1<0$ being antiferromagnetic) in order to provide a dimensionless parametric framework. 
This consideration not only enables one to use the experimental data of exchange couplings, but also paves the way for readers to apply our expressions for similar $\mathrm{Cu}_{4}^{\mathrm{II}}$ complexes with different types of exchange interactions.
Consequently, one finds three different ground states: quantum ferromagnetic $\vert {1,1,1} \rangle_\text{I}$, quantum antiferromagnetic
 $\vert{0,1,1} \rangle$ and classical ferromagnetic $\vert{2,1,1} \rangle$ for $J_2/|J_1|<0$.
The boundary between each two ground states of Eq. (\ref{FourGS}) is given in table \ref{tab:table1}.
For example, the boundary between $\vert {1,1,1} \rangle_\text{I}$ and $\vert{0,1,1} \rangle$ phases is given by $\mu_BB/|J_1| = -\frac{3}{4}K/|J_1|+\frac{1}{2|J_1|}\sqrt{\Lambda}$, while between $\vert {1,1,1} \rangle_\text{I}$ and $\vert {2,1,1} \rangle$ phases is given by $\mu_BB/|J_1|=-1-J_2/|J_1|+\frac{1}{2}K/|J_1|$. The phase diagram pleasantly coincides the magnetization jumps and plateaus of the model under consideration.
All three ground states coexist together at the triple point $K_\text{c}/|J_1|=2J_2/|J_1|=-0.106$ situated nearby the phase boundary between two states $\vert {1,1,1} \rangle_\text{I}$ and $\vert {1,1,1} \rangle_\text{II}$ with the same magnetization value. We discuss this important term in what follows.

In Fig. \ref{GSPT} (b) the ground-state phase diagram of the model is shown in the $\mu_BB/|J_1|-J_2/|J_1|$ plane when $K=0$.
The same ground states to panel Fig.  \ref{GSPT} (a) 
are visible when antiferromagnetic case $J_2/|J_1|<0$ is assumed, while for ferromagnetic range $J_2/|J_1|>0$ a different ground state $\vert {1,1,1} \rangle_\text{II}$ with $M/M_\text{s}=1/2$ appears. Here, a quadruple point is obvious at which four ground states coexist together.

 The most interesting results of the ground-state phase diagram within the $\mu_BB/|J_1|-J_2/|J_1|$ plane is achieved when the model involves with a non zero cyclic four-spin exchange term.  We plot in Figs. \ref{GSPT} (c) and \ref{GSPT} (d) the ground-state phase diagram for the fixed values $K/|J_1|=0.5$ and $K/|J_1|=0.75$, respectively. These figures obviously represent the topology of the first-order phase transitions between the ground states with different magnetization magnitude (solid lines with different colors), and between the ground states  $\vert {1,1,1} \rangle_\text{I}$ and $\vert {1,1,1} \rangle_\text{II}$ with the same magnetization value (vertical dotted line)  that comprise the total magnetization at one-half of saturation magnetization. 
  
 \begin{table*}
\caption{\label{tab:table1} The phase boundaries between four ground states of the considered spin-1/2 Heisenberg model on a tetranuclear square compound (see Fig. \ref{fig:Model}) with Hamiltonian (\ref{Hamitonain}). First column classifies each pair of ground states with the notation 
$\vert S_\text{T}, S_{12}, S_{34}\rangle$ that have mutual boundary. The second column accounts for the eigenvalues of the  $S_\text{T}^z$ operator, while third column stands for the corresponding eigenenergies to the first column. And the last column figures out the relevant boundary curve formula by adopting $|J_1|$ as the energy unit.}
\begin{ruledtabular}
 \begin{tabular}{| m{2cm} | m{1.25cm} | m{6.5cm} | m{5cm} |}
 \centering
Ground states & $S_\text{T}^z$ & Energies &  Boundary curve \\ [0.1cm]
\hline\\ 

 {\centering $\vert {1,1,1} \rangle_\text{I}$}  \newline \newline
 {$\vert 0,1,1 \rangle$}  & 1 \newline\newline\newline\newline 0 & $E_{11}=-\mu_BB+\frac{1}{2}J_1+\frac{1}{2}J_2-\frac{7}{16}K$  \newline \newline $ E_{16}=\frac{1}{2}J_1+\frac{1}{2}J_2+\frac{5}{16}K-\frac{1}{2}\sqrt{\Lambda}$ &
{ $\mu_BB/|J_1|= -\frac{3}{4}K/|J_1|+\frac{1}{2|J_1|}\sqrt{\Lambda}$}
  \\ [0.25cm]
      \hline 
{$\vert {1,1,1} \rangle_\text{I}$}  \newline\newline   {$\vert 2,1,1 \rangle$}    & 1  \newline\newline\newline\newline  2 & $E_{11}=-\mu_BB+\frac{1}{2}J_1+\frac{1}{2}J_2-\frac{7}{16}K$   \newline  $ E_{14}=-2\mu_BB-\frac{1}{2}J_1-\frac{1}{2}J_2+\frac{1}{16}K$ & $\mu_BB/|J_1|=-1-J_2/|J_1|+\frac{1}{2}K/|J_1|$ \\  [0.25cm]
      \hline
{$\vert{1,1,1} \rangle_\text{I}$}  \newline\newline   $\vert {1,1,1} \rangle_\text{II}$    & 1  \newline\newline\newline\newline  -1 & $E_{11}=-\mu_BB+\frac{1}{2}J_1+\frac{1}{2}J_2-\frac{7}{16}K$   \newline  $ E_{14}=-\mu_BB+\frac{1}{2}J_1-\frac{1}{2}J_2-\frac{1}{16}K$ & $J_2/|J_1|=\frac{1}{2}K/|J_1|$ \\  [0.25cm]
\hline
{$\vert {1,1,1} \rangle_\text{II}$}  \newline\newline   {$\vert 2,1,1 \rangle$}    & -1  \newline\newline\newline\newline  2 & $E_{11}=-\mu_BB+\frac{1}{2}J_1-\frac{1}{2}J_2-\frac{1}{16}K$  \newline  $ E_{14}=-2\mu_BB-\frac{1}{2}J_1-\frac{1}{2}J_2+\frac{1}{16}K$ & $\mu_BB/|J_1| = 1$ 
   \\  [0.25cm]
\hline
$\vert {1,1,1} \rangle_\text{II}$ \newline\newline {$\vert 0,1,1 \rangle$}  & -1 \newline\newline\newline\newline 0 & $E_{11}=-\mu_BB+\frac{1}{2}J_1-\frac{1}{2}J_2-\frac{1}{16}K$  \newline $ E_{16}=\frac{1}{2}J_1+\frac{1}{2}J_2+\frac{5}{16}K-\frac{1}{2}\sqrt{\Lambda}$ &
 $ \mu_BB/|J_1|=-J_2/|J_1|-\frac{1}{4}K/|J_1|+\frac{1}{2|J_1|}\sqrt{\Lambda}$
 
\end{tabular}
\end{ruledtabular}
\end{table*}

As a result, we observe that the ground-state phase diagram of the model does not show  phase transition between two ground states with the same magnetization value when the cyclic four-spin exchange term is zero, while for $K/|J_1|>0$ the system exhibits a typical first-order phase transition between them in the ferromagnetic regime $J_2/|J_1|>0$ (see dotted line of Fig. \ref{GSPT} (c)). With increase of the $K/|J_1|$ we witness that the boundary between   $\vert {1,1,1} \rangle_\text{I}$ and $\vert{1,1,1} \rangle_\text{II}$ becomes wider and shifts towards higher values of the coupling constant $J_2/|J_1|$.

\section{Results and discussion}\label{results}
In this section, we discuss the novel results obtained from our  theoretical investigations for the effects of the cyclic four-spin interaction on the geometric $\Pi_4$ average of tangles as a measure of the tetrapartite quantum entanglement. Next, we examine the whole entanglement $\Pi_4$, bipartite negativity $N_\text{AB}$ as a measure of pairwise quantum entanglement, average fidelity $F_\text{A}$ and QFI of the tetranuclear $\mathrm{Cu}^{\mathrm{II}}$ square complex $[\text{Cu}_4\text{L}_4(\text{H}_2\text{O})_4](\text{ClO}_4)_4$ at temperatures as high as the room temperature scale and even more. This compound exceptionally offers an experimental prototype with robust quantum correlations allowed us to pleasantly elucidate a theoretical insight into the quantum entanglement and the power of teleportation of a four-qubit quantum system with strong antiferromagnetic Heisenberg interaction at sufficiently high temperatures. Some interesting remarks will be delivered when  a typical cyclic four-spin interaction is considered for this exampled model.

\subsection{Theoretical investigation of the tetrapartite entanglement}
For our favorite system, there are different tetrapartite entanglement quantifiers. Among them, $\pi-$tangle and four-tangle can be employed as the most generic ones \cite{Li2016,Dong2019,Arenas2019,Dong2020}. 
To extend the definition of the tetrapartite entanglement we invoke the whole entanglement measure ${\Pi}_4$. 
For a given four-qubit state we begin by defining below negativities

\begin{equation}\label{negativities}
\begin{array}{lcl}
N_\mathrm{A(BCD)}=\vert\vert\rho ^{T_\mathrm{A}}_\mathrm{ABCD}\vert\vert -1,
\\[0.1cm]
N_\mathrm{AB(CD)}=\vert\vert\rho ^{T_\mathrm{AB}}_\mathrm{ABCD}\vert\vert -1,
\\[0.1cm]
N_\mathrm{A(BC)}=\vert\vert\rho ^{T_\mathrm{A}}_\mathrm{ABC}\vert\vert -1,
\\[0.1cm]
N_\mathrm{AB}=\vert\vert\rho ^{T_\mathrm{A}}_\mathrm{AB}\vert\vert -1,
\end{array}
\end{equation}
in which $\rho_\text{ABCD}(T)$ $=\exp \left( -\beta {H}\right) /Z$ denotes the density matrix of four-spin system with Hamiltonian (\ref{Hamitonain}) in thermal equilibrium at temperature $T$ where $Z=\mathrm{Tr}\left[ \exp(-\beta {H})\right] $ denotes the partition function of the four-spin system and $\beta ={1/{{k_B}T}}$.
The density operator $\rho_\text{ABCD}(T)$ can be thus defined as
\begin{equation} 
\rho_\mathrm{ABCD} (T)=\frac{1}{Z}\sum\limits_{l=1}^{16}\exp \left( -\beta E_{l}\right)
\left\vert \psi _{l}\right\rangle \left\langle \psi _{l}\right\vert ,
\label{state}
\end{equation}%
where $E_{l}$ and $\left\vert \psi _{l}\right\rangle $ are, respectively, eigenvalues and eigenvectors of the Hamiltonian $H$.
The nonzero coefficients of the density matrix $\rho_\mathrm{ABCD}$ are given in Appendix \ref{sec.appendixA}.
 ${T_\mathrm{A}}$ and ${T_\mathrm{AB}}$ indicate the partial transpose over the part $\mathrm{A}$ and pair $\mathrm{AB}$, respectively, and $\vert\vert \cdot \vert\vert$ stands for the trace norm of a matrix represented as  $\vert\vert O \vert\vert= \mathrm{Tr} {\sqrt {O^{\dagger}O}}$ \cite{Horn1985}.  $N_{\mathrm{A(BCD)}}, N_{\mathrm{AB(CD)}}, N_{\mathrm{A(BC)}}$, and $N_{\mathrm{AB}}$ describe the entanglement between two parts, such as $1-3$ tangle, $2-2$ tangle, $1-2$ tangle, and $1-1$ tangle entanglement. For instance, $N_{\mathrm{A(BCD)}}$ describes the entanglement between the part $\mathrm{A}$ and the others $\{\mathrm{B},\mathrm{C},\mathrm{D}\}$. Similarly,
  $N_{\mathrm{AB(CD)}}$ means the negativity between pair $\{\mathrm{A},\mathrm{B}\}$ and remaining pair $\{\mathrm{C},\mathrm{D}\}$, while
   $N_{\mathrm{AB}}$ introduces the bipartite negativity between part $\mathrm{A}$ and part $\mathrm{B}$ where  $\rho _{\mathrm{AB}}=\mathrm{Tr}_{\mathrm{CD}}[\rho_{\mathrm{ABCD}}]$. 
 $N_{\mathrm{A(BC)}}$ demonstrates the entanglement negativity between part $\mathrm{A}$ and pair $\{\mathrm{B},\mathrm{C}\}$ after tracing over $\mathrm{D}$. 
 Here, the  $1-1$ and $1-3$ tangles satisfy the following Coffman–Kundu–Wootters (CKW) monogamously inequality relation \cite{Wootters2000}

\begin{equation}\label{ABCD}
\begin{array}{lcl}
N^2_{\mathrm{A(BCD)}}\geq N^2_{\mathrm{AB}}+N^2_{\mathrm{AC}}+N^2_{\mathrm{AD}}.
\end{array}
\end{equation}
Accordingly,  four-tangle entanglement (residual tangles) can be characterized as

\begin{equation}\label{abcd}
\begin{array}{lcl}
\pi_\mathrm{A}=N^2_{\mathrm{A(BCD)}}-N^2_{\mathrm{AB}}-N^2_{\mathrm{AC}}-N^2_{\mathrm{AD}},
\\
\pi_\mathrm{B}=N^2_{\mathrm{B(ACD)}}-N^2_{\mathrm{BA}}-N^2_{\mathrm{BC}}-N^2_{\mathrm{BD}},
\\
\pi_\mathrm{C}=N^2_{\mathrm{C(ABD)}}-N^2_{\mathrm{CA}}-N^2_{\mathrm{CB}}-N^2_\mathrm{CD},
\\
\pi_\mathrm{D}=N^2_\mathrm{D(ABC)}-N^2_\mathrm{DA}-N^2_\mathrm{DB}-N^2_\mathrm{DC}.
\end{array}
\end{equation}
In the one hand, the existence of four different residual tangles $\pi_\mathrm{A}$, $\pi_\mathrm{B}$, $\pi_\mathrm{C}$ and $\pi_\mathrm{D}$ encourages one to use $\pi_4-$tangle; $\pi_4=\frac{1}{4}(\pi_\mathrm{A}+\pi_\mathrm{B}+\pi_\mathrm{C}+\pi_\mathrm{D})$ for measuring the whole degree of entanglement of the tetrapartite spin system.
 On the other hand, another measure of whole entanglement, described as geometric mean ${\Pi}_4$ \cite{Alcaine2008}, can be defined as

\begin{equation}\label{CD}
\begin{array}{lcl}
{\Pi}_4=\sqrt[4]{\pi_\mathrm{A}\pi_\mathrm{B}\pi_\mathrm{C}\pi_\mathrm{D}}.
\end{array}
\end{equation}
In what follows, we will examine the whole entanglement of the tetrameric spin-1/2 Heisenberg system (\ref{Hamitonain}) by using ${\Pi}_4$.

\begin{figure*}[t]
\centering
\includegraphics[scale=0.5,trim=20 00 20 20, clip]{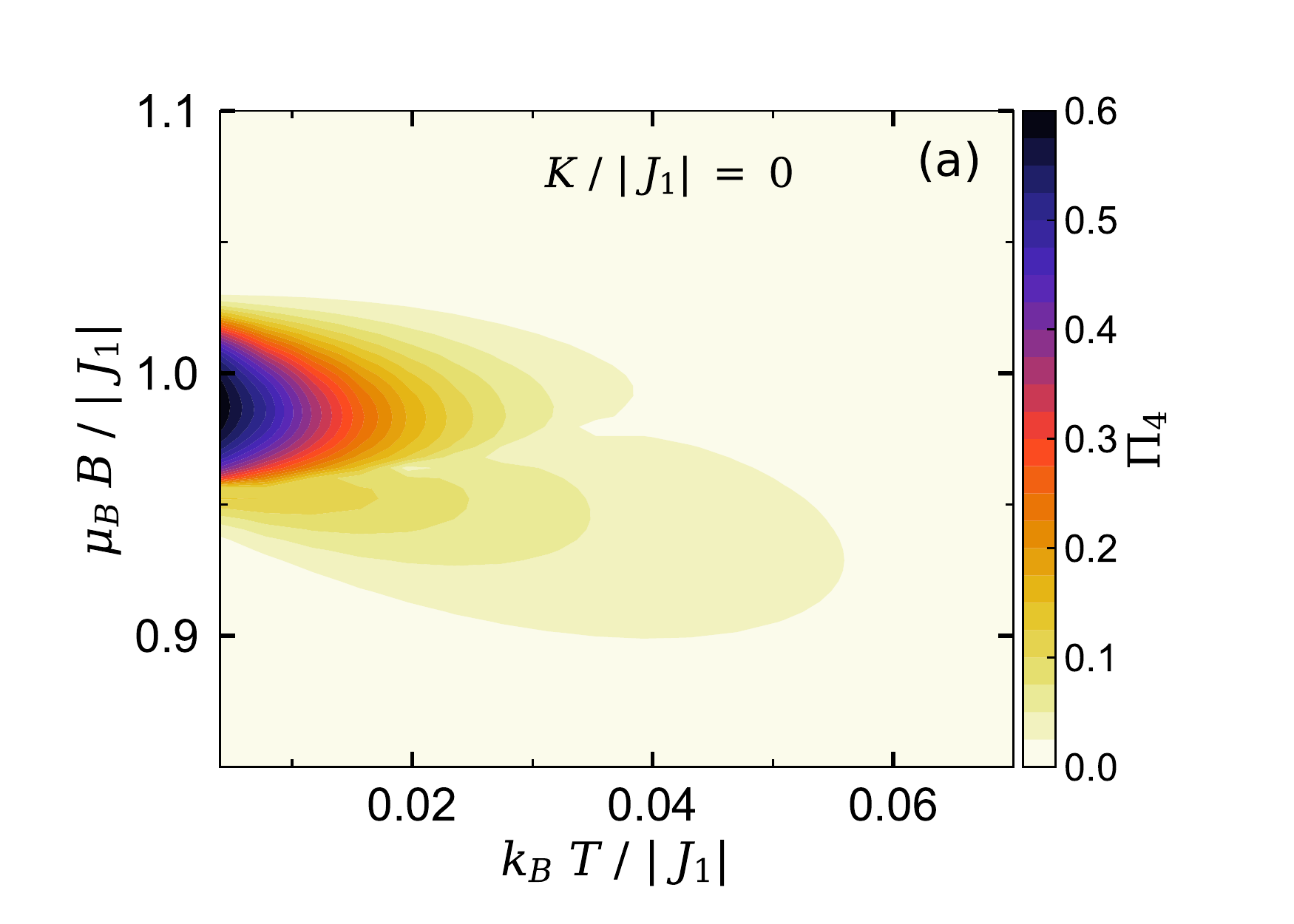}
\includegraphics[scale=0.5,trim=20 00 20 20, clip]{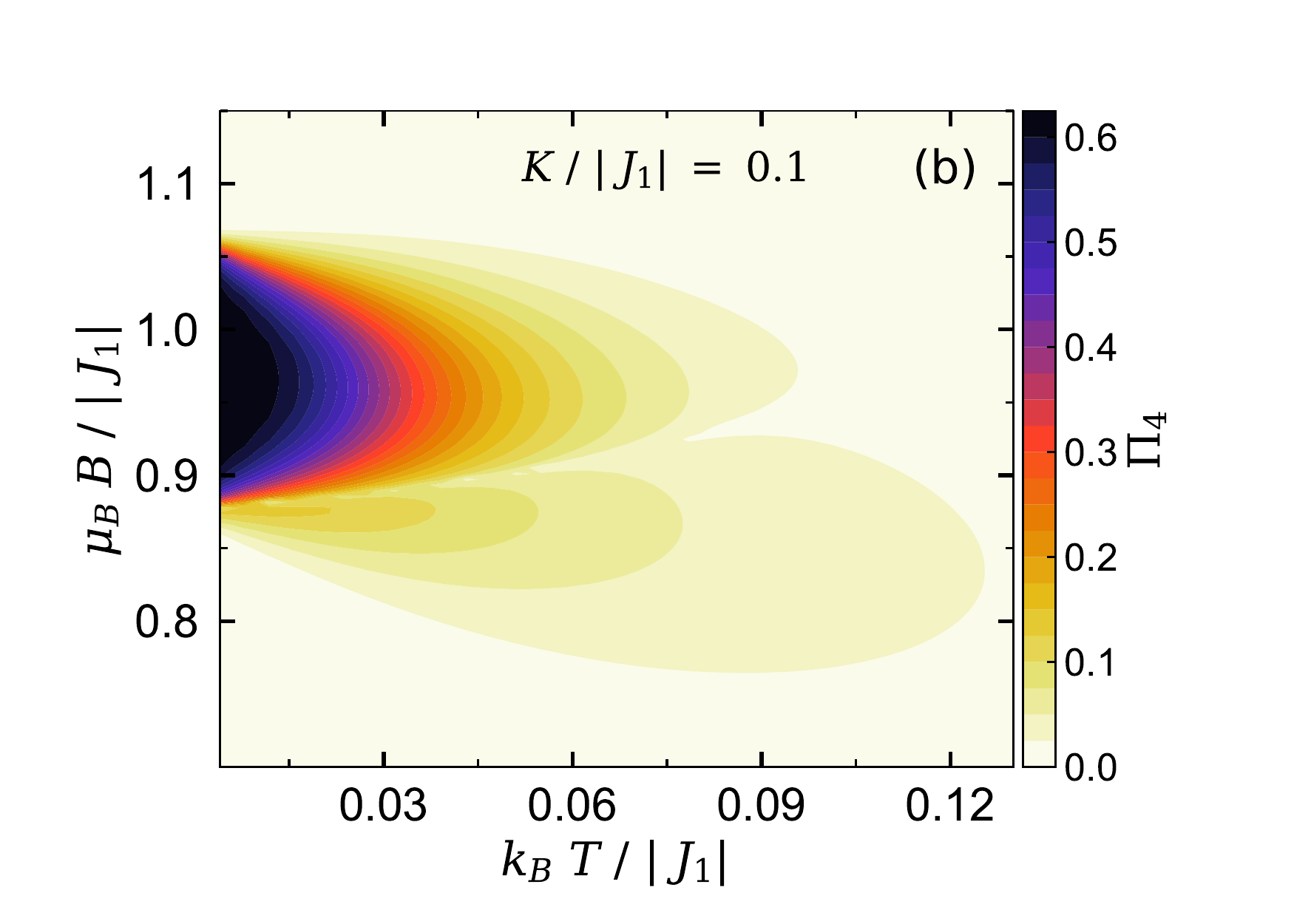} %
\includegraphics[scale=0.5,trim=20 00 20 30, clip]{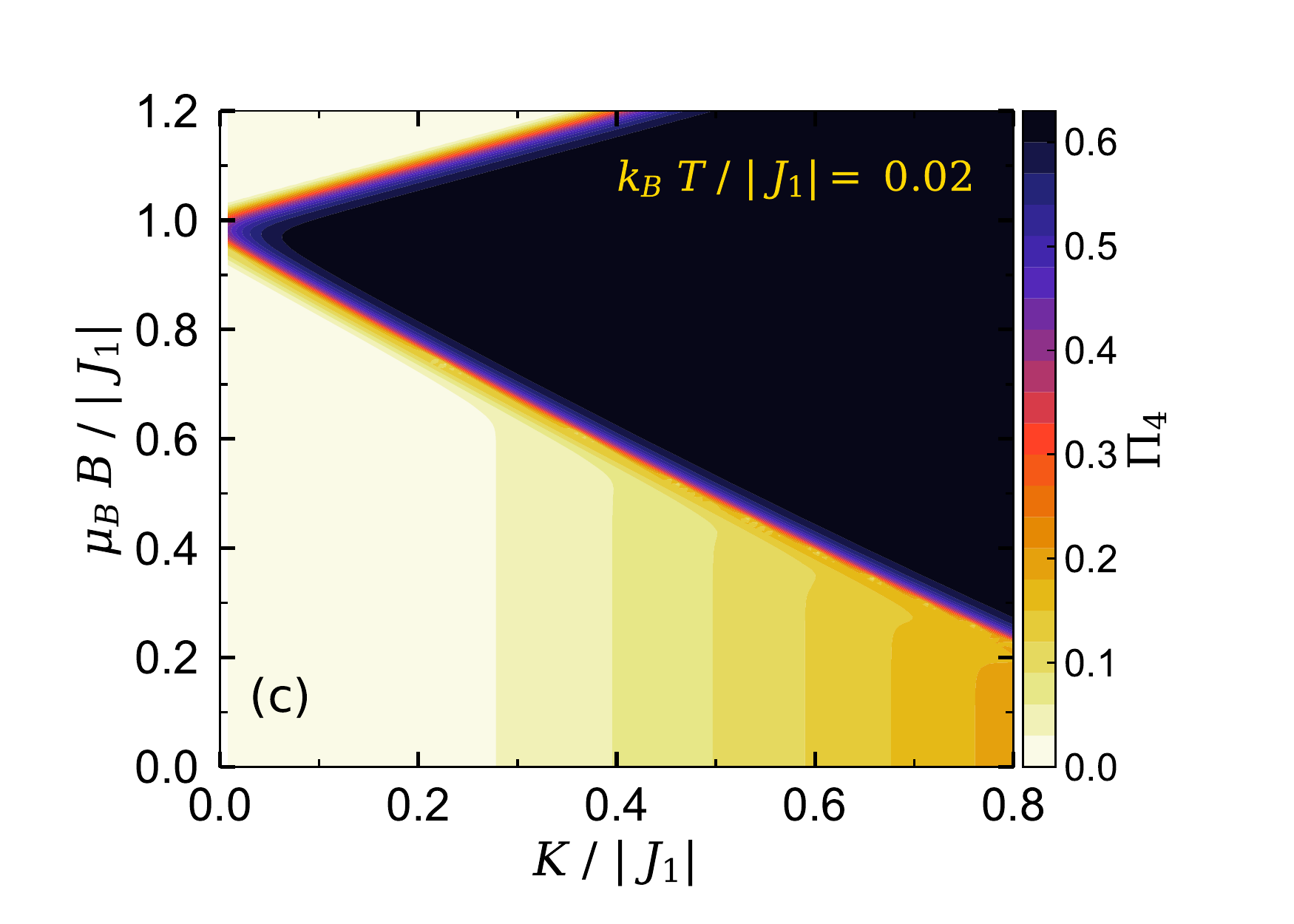}
\includegraphics[scale=0.5,trim=20 00 00 30, clip]{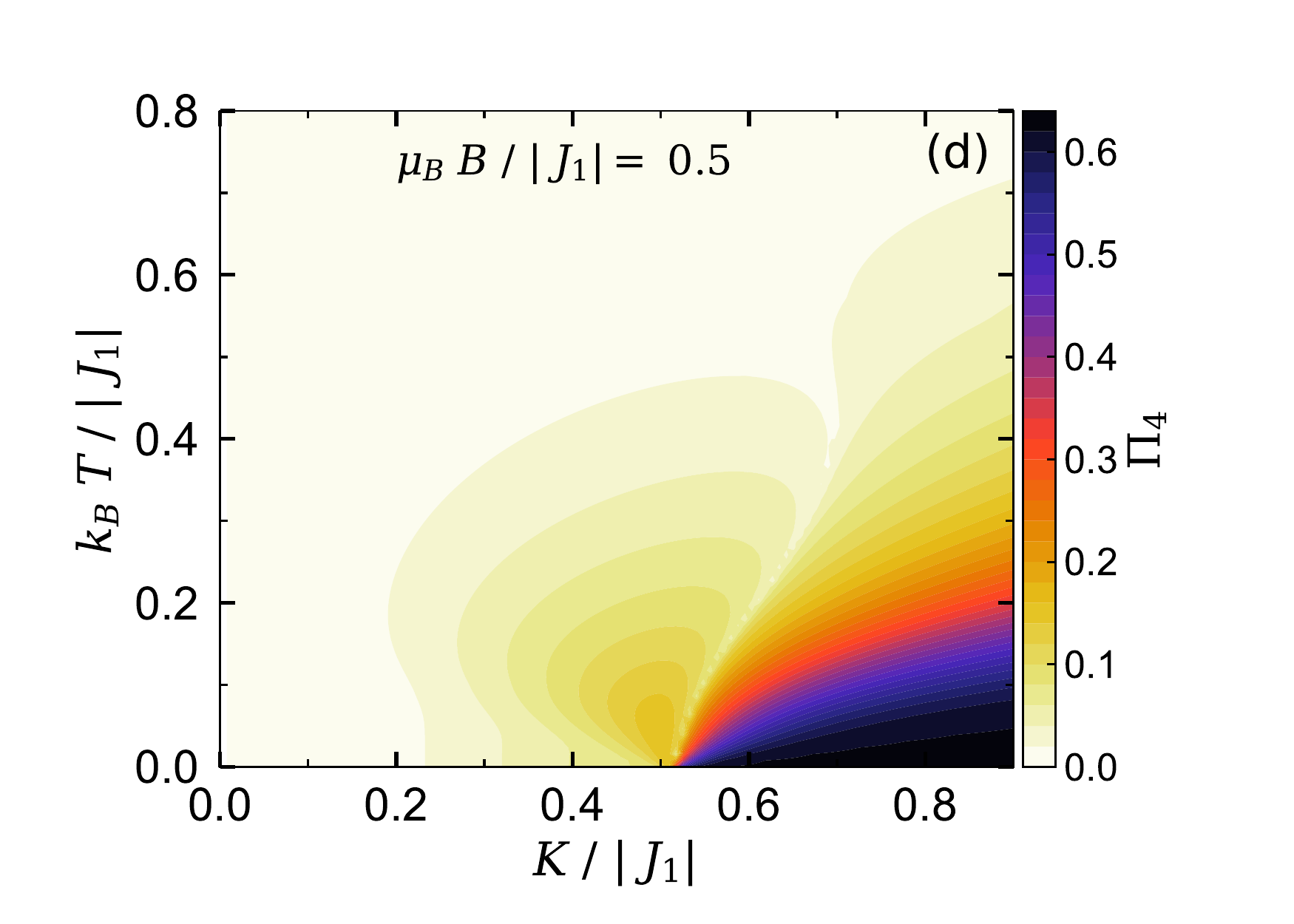} %

\caption{Contour plot of the whole entanglement negativity $\Pi_4$ of the spin-1/2 antiferromagnetic Heisenberg XXX model on a tetranuclear square compound with Hamiltonian (\ref{Hamitonain}) consisting of the cyclic four-spin interaction $K/|J_1|$. Panels (a) and (b) show the whole entanglement negativity $\Pi_4$ in the $\mu_BB/|J_1|-k_BT/|J_1|$ plane for $K/|J_1|=0$ and $K/|J_1|=0.1$, respectively. (c) Contour plot of $\Pi_4$ in the $\mu_BB/|J_1|-K/|J_1|$ plane at low temperature $k_BT/|J_1|=0.02$. (d)  The whole entanglement $\Pi_4$ in the $k_BT/|J_1|-K/|J_1|$  plane at $\mu_BB/|J_1|=0.5$. According to the coordinates of the triple point appeared in the ground-state phase diagram of the model (see Fig. \ref{GSPT}(a)), weak antiferromagnetic exchange coupling $J_2/|J_1|=-0.053$ has been assumed.
}
\label{Pi_4}
\end{figure*}

We represent in Fig. \ref{Pi_4} (a) the whole entanglement $\Pi_4$ of the four-qubit state $\rho_\mathrm{ABCD}$ corresponds to the  spin-1/2 Heisenberg model on a square-shaped complex in the $\mu_BB/|J_1|-k_BT/|J_1|$ plane when $K/|J_1|=0$. 
A weak antiferromagnetic exchange coupling $J_2/|J_1|=-0.053$ is considered that coincides the coordinates of the triple point at which three ground states coexist together (see Fig.  \ref{GSPT}(a)).
It is crystal clear that the tetrapartite entanglement  $\Pi_4$ is in its maximum value $\Pi_4\approx 0.621$ at sufficiently low temperature $k_BT/|J_1|<0.01$  when the magnetic field is as large as the absolute value $|J_1|$. 
In Fig. \ref{Pi_4} (b) we show the whole entanglement in the $\mu_BB/|J_1|-k_BT/|J_1|$ plane when a small nonzero value $K/|J_1|=0.1$ is considered. One sees that a nonzero positive value of the cyclic four-spin exchange $K/|J_1|$ remarkably broadens the region at which the tetrapartite entanglement exists.
Figure \ref{Pi_4} (c) displays the tetrapartite entanglement negativity in the $\mu_BB/|J_1|-K/|J_1|$ plane at low temperature $k_BT/|J_1|=0.02$.
This figure practically reproduces the ground-state phase diagram plotted in Fig. \ref{GSPT} (a). The effect of cyclic four-spin exchange on the whole entanglement negativity $\Pi_4$ is obvious in this figure. 
When the ring exchange $K/|J_1|$ increases the maximum value of the whole entanglement can be achieved at lower magnetic fields.
We display in Fig. \ref{Pi_4} (d) the function $\Pi_4$ in the  $k_BT/|J_1|-K/|J_1|$ plane when the magnetic field is fixed at $\mu_BB/|J_1|=0.5$.
From this figure one finds that with increase of the temperature the both entangled states  
$\vert {0,1,1} \rangle$ and  $\vert {1,1,1} \rangle_\text{I}$
lose their entanglement degree. The phase boundary between these states are evident by a narrow pale stripe. This boundary separates two phases even at  high enough temperature region ($k_BT/|J_1|\approx 0.5$). It is also interesting to notify that this separation starts from $K/|J_1|\approx \mu_BB/|J_1|$. It could be concluded that the whole entanglement in the model under consideration is still sizable at sufficiently high temperature when a special cyclic four-spin interaction comes into play.

\subsection{Experimental testing of theoretical results}
From now on, we concentrate on the particular case $[\text{Cu}_4\text{L}_4(\text{H}_2\text{O})_4](\text{ClO}_4)_4$ that showcasts a good experimental platform of a strong antiferromagnetic Heisenberg XXX model on a square-shaped compound. Then, we theoretically investigate the effects of a deemed cyclic four-spin interaction on the whole entanglement negativity, bipartite entanglement negativity, the ability of quantum teleportation and QFI of this experimental example by considering actual values of the exchange couplings $J_1/k_B=-638\; \text{K}$ and $J_2/k_B=-34\; \text{K}$ reported in Ref. \cite{Giri2011}.

The spin density of the complex $[\text{Cu}_4\text{L}_4(\text{H}_2\text{O})_4](\text{ClO}_4)_4$ is mainly distributed between the $d_{x^2-y^2}$ orbital of the copper atoms and the hybrid orbitals of the donor atoms in the equatorial plane. Hence, it is expected that this compound does not consists of actual anisotropy neither two-ion exchange anisotropy nor Dzyaloshinsky--Morya (DM) interaction due to its geometry.
 For that reason, we proceeded to theoretically examine the effect size value of the cyclic four-spin interaction $K/k_B$ on the temperature dependence of $\chi_\text{M}T$ product of the complex $[\text{Cu}_4\text{L}_4(\text{H}_2\text{O})_4](\text{ClO}_4)_4$, and we eventually realized that it is much smaller than other anisotropy parameters specially for the temperature range $T < 100\ \text{K}$. 
 The results concerning these findings have not been added to this work for brevity.
Hopefully, the theoretical studies reported in the current work will stimulate
future experimental testing of the entanglement negativity, quantum teleportation and QFI of the tetranuclear square complex $[\text{Cu}_4\text{L}_4(\text{H}_2\text{O})_4](\text{ClO}_4)_4$ at high temperature through modern experimental procedures.

\subsubsection{Tetrapartite entanglement}

\begin{figure}[tbp]
\centering
\includegraphics[width=8.25cm, height=6cm]{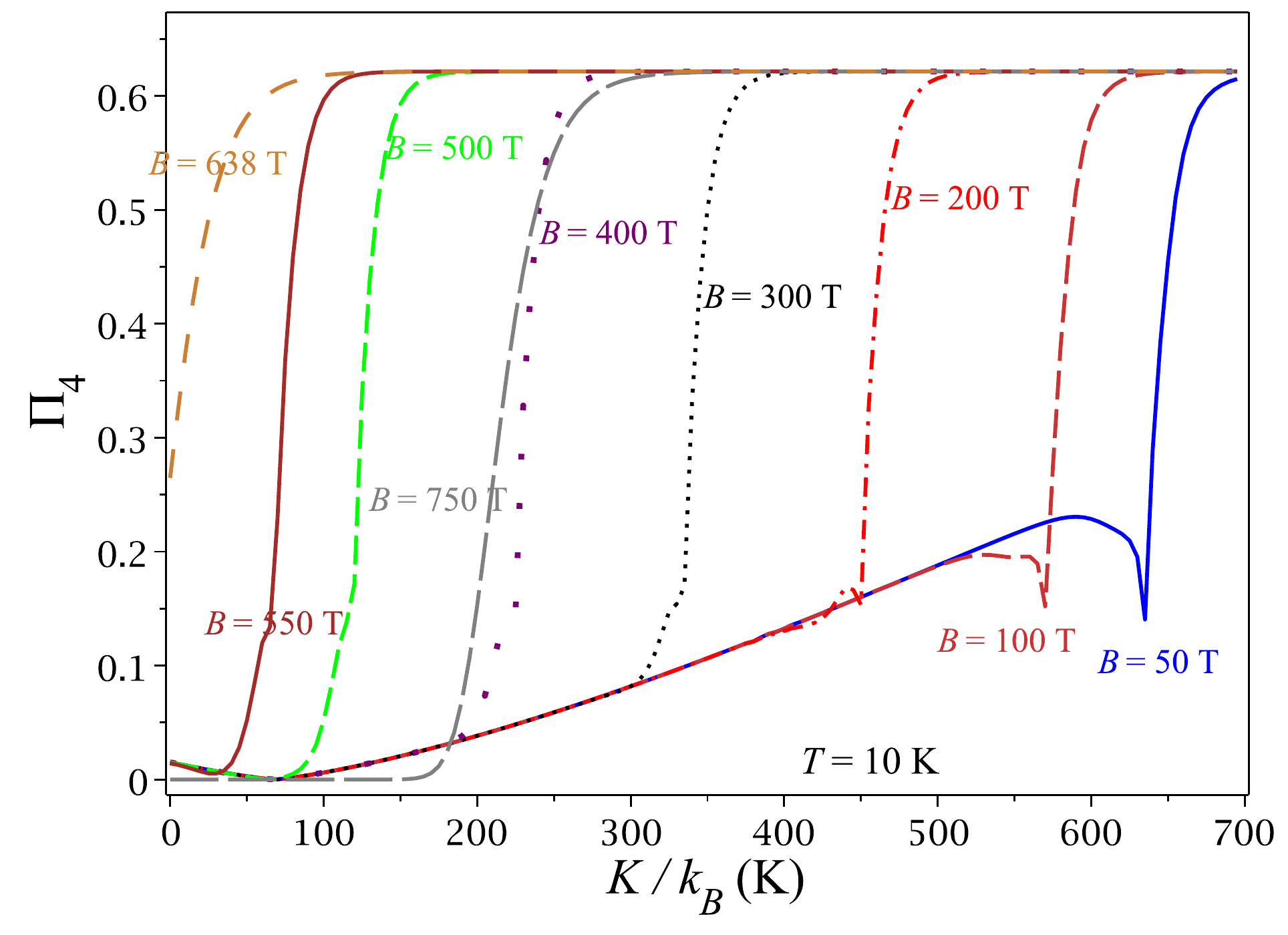} %
\caption{ Tetrapartite entanglement $\Pi_4$ of the complex $[\text{Cu}_4\text{L}_4(\text{H}_2\text{O})_4](\text{ClO}_4)_4$ as a function of the assumed cyclic four-spin interaction $K/k_B$ for a few selected values of the magnetic field at moderate  temperature $T=10\; \text{K}$ such that $J_1/k_B=-638\; \text{K}$, $J_2/k_B=-34\; \text{K}$.}
\label{PI_K}
\end{figure}
The cyclic four-spin exchange dependence of the geometric $\Pi_4$ average is manifested in Fig. \ref{PI_K} at moderate temperature $T=10\; \text{K}$ for different values of the magnetic field where exchange couplings are taken up as actual values $J_1/k_B=-638\; \text{K}$ and $J_2/k_B=-34\; \text{K}$.
It is evident from Fig. \ref{PI_K} that for the magnetic fields smaller than absolute value $|J_1|$ the whole entanglement $\Pi_4$ of the
 tetranuclear Cu$^\text{II}_4$ complex exhibits an unconventional minimum at $K/k_B=68\; \text{K}$ such that $\Pi_4 = 0$.
 It is worthwhile to note that a very small value of the cyclic four-spin interaction ($K/k_B\ll |J_1|$) can still alters the degree of the whole entanglement negativity $\Pi_4$ for $B\ll |J_1|$ (see blue solid line of Fig. \ref{PI_K}). This finding is in accordance with that of discussed in Fig. \ref{Pi_4}(b).
  When $K$ increases further than the observed minimum point ($K/k_B=68\; \text{K}$), the entanglement degree of the system is significantly enhanced. We choose the intriguing value $K/k_B=68\; \text{K}$ then examine the bipartite entanglement negativity $N_\mathrm{AB}$ in what follows. 
In addition, a reentrant behavior can be observed in $\Pi_4$ curve close to the critical points at which phase transition occurs between $\vert 0,1,1 \rangle$ and $\vert 1,1,1\rangle_\text{I}$ states (see Figs. \ref{GSPT} (a) and \ref{Pi_4}(c)).
On the other hand, when magnetic fields greater than the absolute value $|J_1|$ are considered the whole entanglement $\Pi_4$ starts to increase from a critical point and rapidly reaches its maximum value.

\subsubsection{Bipartite entanglement}
In this part we will put forward a theoretical inspection for the bipartite entanglement between spins A and B of the complex $[\text{Cu}_4\text{L}_4(\text{H}_2\text{O})_4](\text{ClO}_4)_4$ with reduced density matrix $\rho _\mathrm{AB}$ (see Eq. (\ref{state2}) and Appendix \ref{sec.appendixA}).
 Since the system with Hamiltonian (\ref{Hamitonain}) is symmetric, all of the reduced density 
matrices have the same formation, i.e., they are block diagonal. Therefore, In the standard basis $\{\left\vert 00\right\rangle
,\left\vert 01\right\rangle ,\left\vert 10\right\rangle ,\left\vert11\right\rangle \}$ the reduced density matrix 
$\rho _\mathrm{AB}(T)=\mathrm{Tr}_\mathrm{CD}\left[\rho_\mathrm{ABCD} (T)\right] $ is given by%
\begin{equation}
\rho _\mathrm{AB}(T)=%
\begin{pmatrix}
\varrho _{11} & 0 & 0 & 0 \\
0 & \varrho _{22} & \varrho _{23} & 0 \\
0 & \varrho^{*}_{23} & \varrho_{33} & 0 \\
0 & 0 & 0 & \varrho_{44}%
\end{pmatrix}%
 \label{state2}
\end{equation}
where $\varrho^{*}_{23}=\varrho_{23}$. Nonzero elements of the above density operator are directly connected to the coefficients of the total density operator $\rho _\mathrm{ABCD}(T)$ as we have composed in Appendix \ref{sec.appendixA}.
In order to investigate bipartite thermal entanglement of the reduced density matrix $\rho_\mathrm{AB}$, we employ negativity $N_\mathrm{AB}$ (\ref{negativities}) that is one the most widely accepted  measures of entanglement for a two-qubit system. The case $N_\mathrm{AB}=0$ happens when the system state is separable,
 whereas $N_\mathrm{AB}=1$ indicates maximally entangled state. 

\begin{figure*}[tbp]
\centering
\includegraphics[width=8.25cm, height=5.75cm]{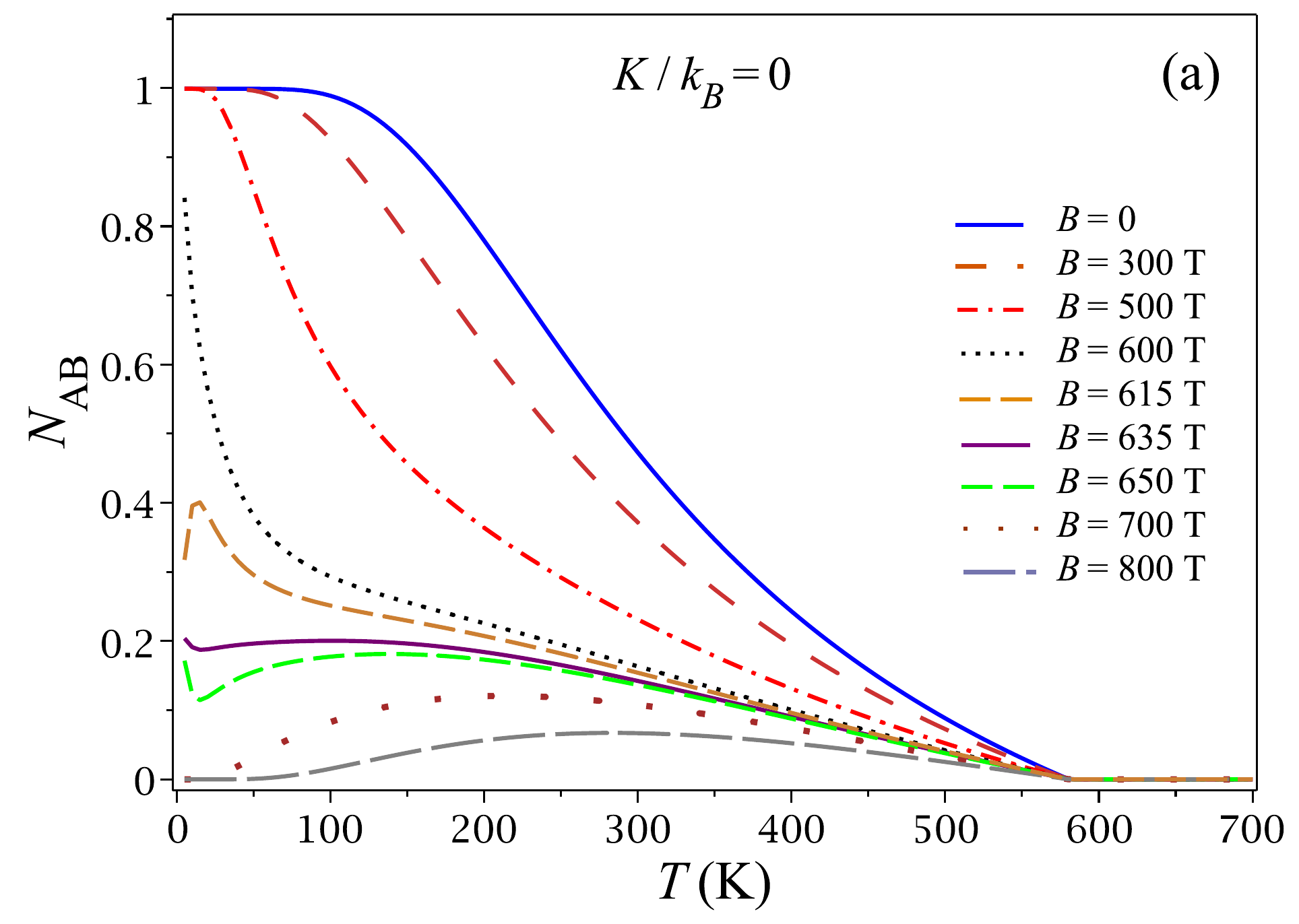} %
\includegraphics[width=8cm, height=5.75cm]{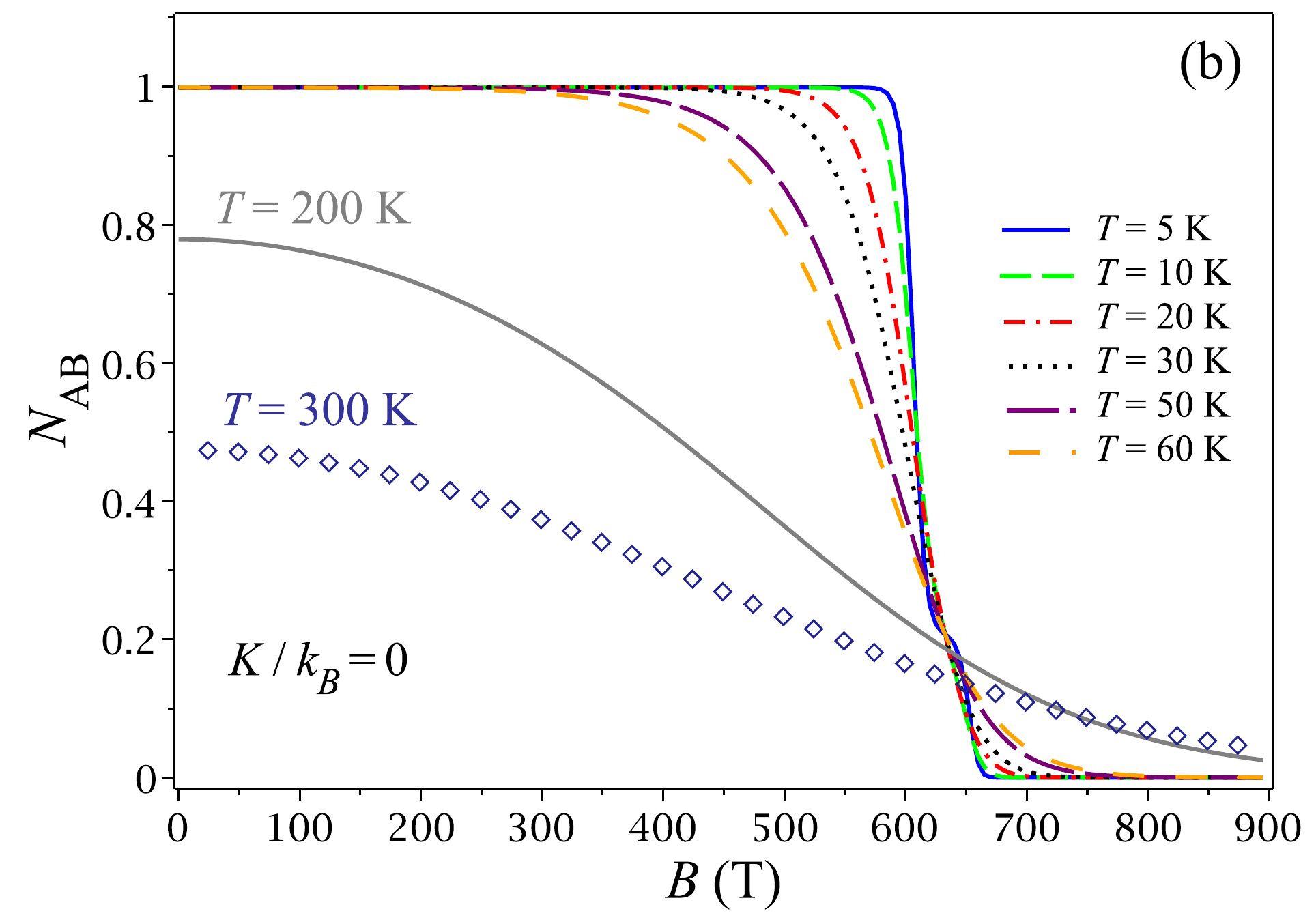} %

\caption{(a) Bipartite entanglement negativity $N_\mathrm{AB}$ versus temperature for several fixed values of the magnetic field when  $K=0$, taking $J_1/k_B=-638\; \text{K}$ and $J_2/k_B=-34\; \text{K}$.
   (b) Negativity $N_\mathrm{AB}$ as a function of magnetic field for several selected temperatures and the same set of exchange couplings to panel (a).}
\label{con}
\end{figure*}

We have shown in Fig. \ref{con} (a) the bipartite negativity $N_\mathrm{AB}$ as a function of the temperature $T$, for several values of the magnetic field $B$ at fixed exchange interactions $J_1/k_B=-638\; \text{K}$ and $J_2/k_B=-34\; \text{K}$, where the cyclic four-spin interaction $K$ is zero. It is clear that the negativity $N_\mathrm{AB}$  decreases  from its maximum value $N_\mathrm{AB}=1$ when $T$ and $B$ increase. 
We find maximum entanglement for notably high temperature ($T\approx 100\; \text{K}$) and high magnetic fields ($B< 300\; \text{T}$). This quantity remains robustly alive for temperatures higher than room temperature. Interestingly, there is an entanglement death at a critical temperature $T_\text{c}\approx 582\; \text{K}$ that is comparable with the absolute value of stronger antiferromagnetic coupling constant $|J_1|$ between spins A and B.

The bipartite entanglement negativity $N_\mathrm{AB}$ as a function of the magnetic field is displayed in Fig. \ref{con} (b) for several temperatures and the same exchange couplings to panel Fig. \ref{con} (a). 
One observes maximum pairwise entanglement at low temperature and remarkably high magnetic fields with comparable magnitude to 
$|J_1|$.
Close to $B=|J_1|$ there is a sharp decrease in $N_\mathrm{AB}$ curve until a narrow plateau is observed at $N_\mathrm{AB}\approx 0.207$, then it decreases and rapidly vanishes at a critical magnetic field $B>|J_1|$. It is evident that nearby room temperature the pairwise entanglement $N_\mathrm{AB}$ in this model is sizable even for strong magnetic fields (see solid gray line and diamonds).

\begin{figure*}[tbp]
\centering
\includegraphics[width=8cm, height=5.75cm]{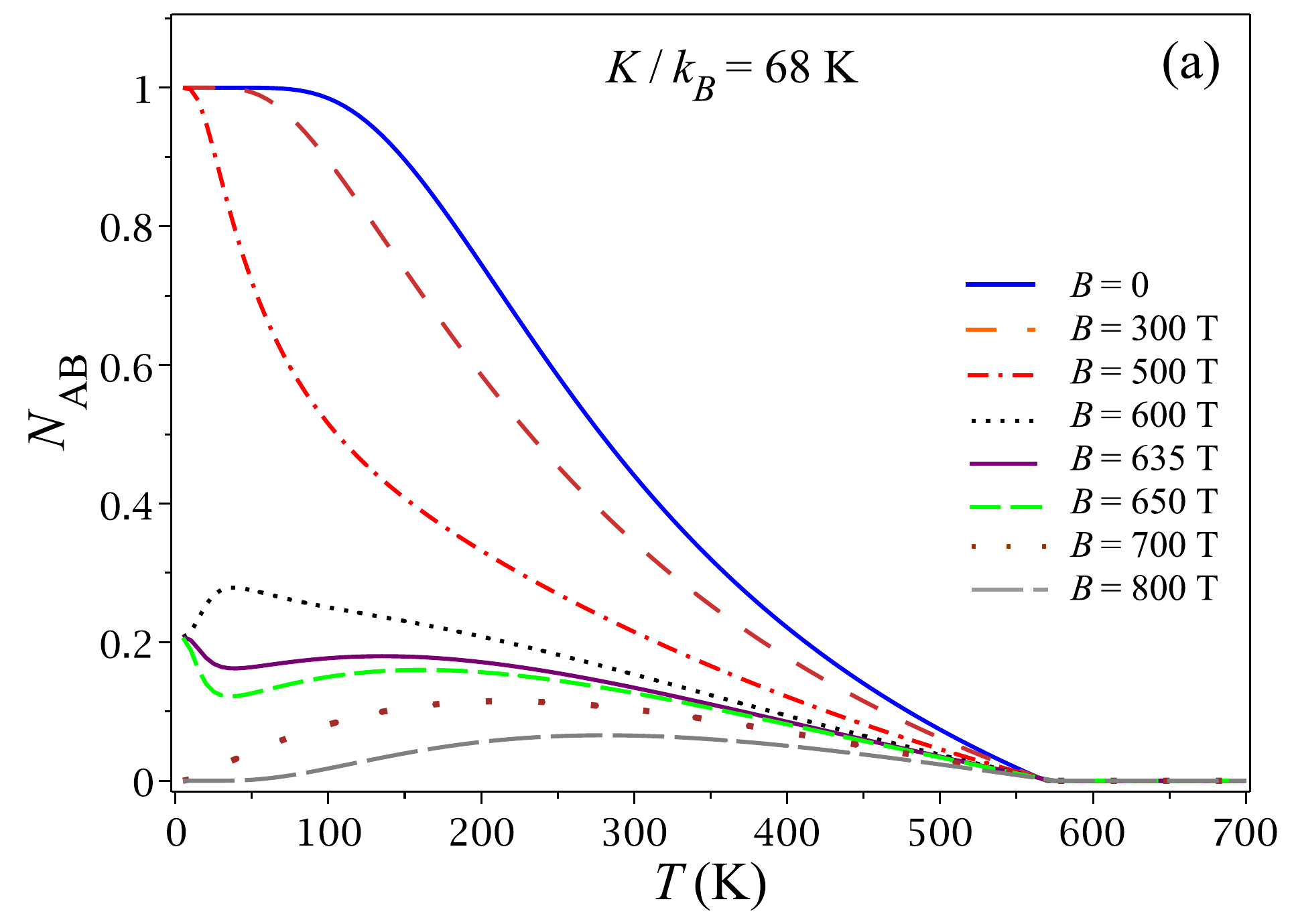} %
\includegraphics[width=8cm, height=5.75cm]{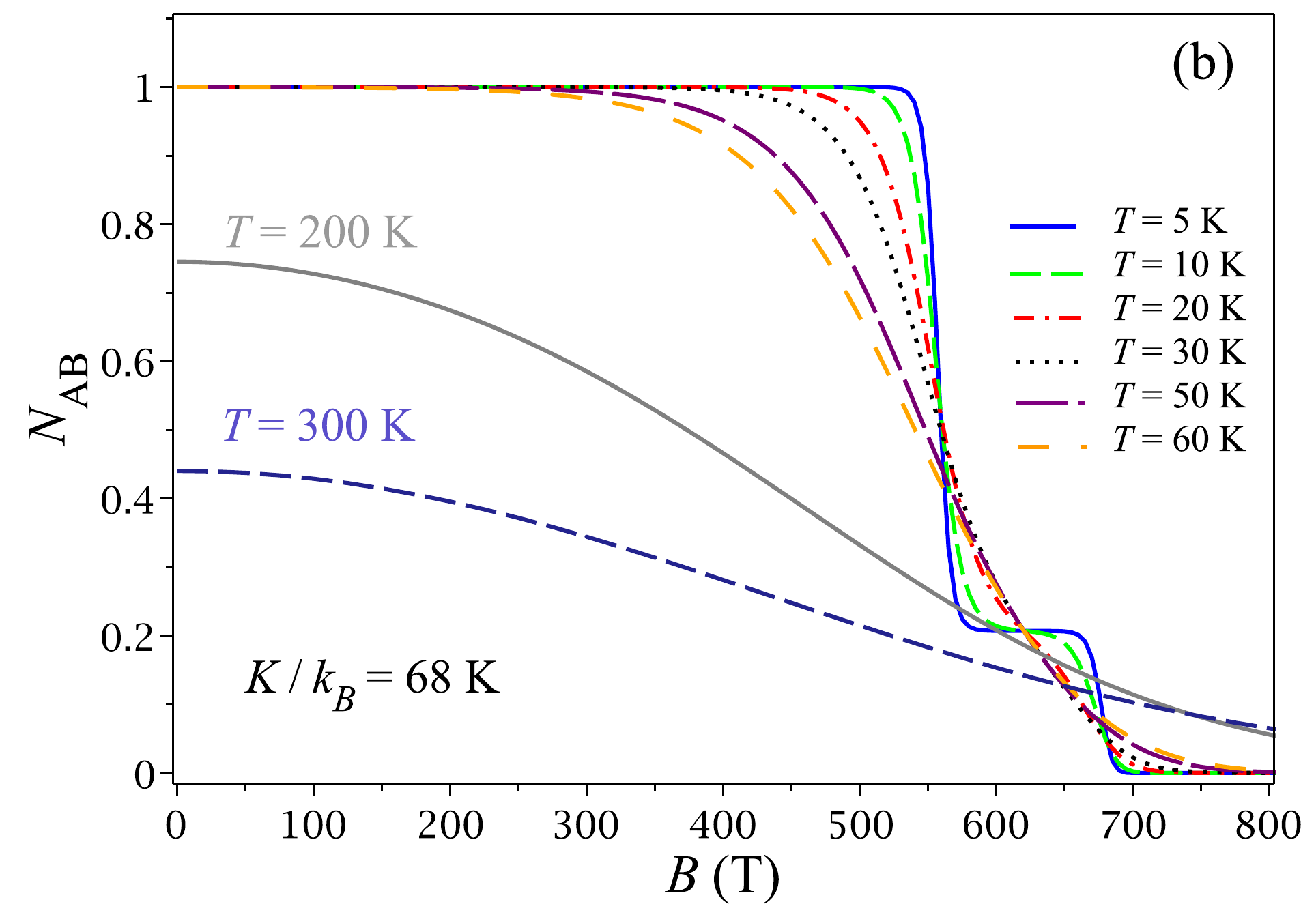} %

\caption{(a) Bipartite entanglement negativity $N_\mathrm{AB}$ versus temperature for several fixed values of the magnetic field, taking $J_1/k_B=-638\; \text{K}$ and $J_2/k_B=-34\; \mathrm{K}$, when $K/k_B= 68\; \mathrm{K}$.
   (b) Magnetic field dependence of the bipartite entanglement $N_\mathrm{AB}$ for  various fixed values of the temperature, assuming the same set of exchange couplings to panel (a).
}
\label{conK50}
\end{figure*}

By inspecting Fig. \ref{conK50} (a) we see that when a nonzero value of the cyclic four-spin term, $K/k_B=68\; \text{K}$, is assumed the entanglement degree decreases for the magnetic fields as large as $|J_1|$ (see dotted line).
Figure \ref{conK50} (b) illustrates the field dependence of the bipartite negativity $N_\mathrm{AB}$ for various selected values of the temperature and fixed $K/k_B=68\; \text{K}$ that is very small compared with the $|J_1|$. 
By Comparing solid gray lines and diamonds plotted in Figs. \ref{con} (b) and \ref{conK50} (b), one quickly realizes that the interplay of even a small cyclic four-spin interaction in the model leads to weaken the pairwise entanglement between Cu atoms interacted through the stronger antiferromagnetic exchange coupling $J_1$ nearby the room temperature. Also, the middle step appeared in the negativity curve becomes wider when a typical cyclic exchange interaction $K>0$ is considered.
 Referring to Fig. \ref{GSPT} (a), one sees that emerged step is accompanied with the ground state $\vert 1,1,1\rangle_\text{I}$ and the change in the $N_\mathrm{AB}$ behavior is in a good agreement with the ground-state phase transition of the model in $B-K$ plane.  When the bipartite entanglement $N_\mathrm{AB}$ is maximum, the system's state is entangled state $\vert 0,1,1 \rangle$, whereas for the case $N_\mathrm{AB}=0$ the system is in fully polarized state $\vert 2,1,1 \rangle$.

\begin{figure}[tbp]
\centering
\includegraphics[width=8cm, height=5.75cm]{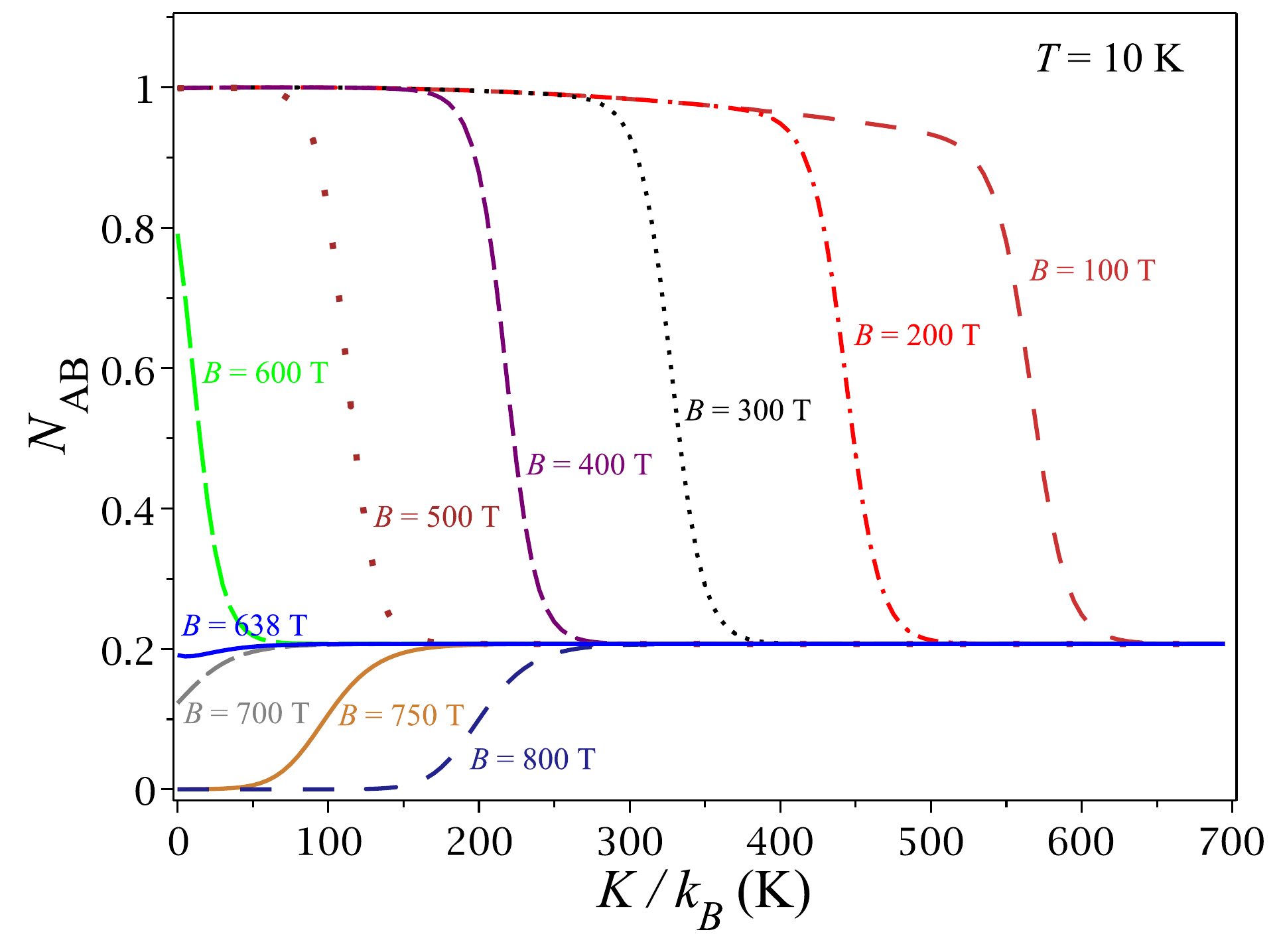} %

\caption{ Bipartite entanglement negativity as a function of the cyclic four-spin interaction $K$ for a few selected values of the magnetic field and fixed  temperature $T=10\; \text{K}$ such that $J_1/k_B=-638\; \text{K}$, $J_2/k_B=-34\; \text{K}$.}
\label{conK}
\end{figure}

The negativity $N_\mathrm{AB}$ behavior versus the ring exchange $K$ is depicted in Fig. \ref{conK} for a few selected values of
the magnetic field at  $T=10\; \text{K}$. 
Principally, this quantity behaves  different from the $\Pi_4$ against the cyclic four-spin exchange $K$.
As can be seen in this figure, $N_\mathrm{AB}$ remains almost steady at 
 value $N_\mathrm{AB} \approx 0.207$ for the magnetic field $B\approx |J_1|$ (see blue solid line).
For magnetic fields smaller than this value, when $K$ increases monotonically the negativity $N_\mathrm{AB}$ decreases from its maximum value and drops down to approximately one-third of maximum value of the whole entanglement $\Pi_4^\mathrm{max}\approx 0.621$, namely $N_\mathrm{AB} \approx 0.207$, at a specific point that its position on the $K-$axis can be controlled by the strength of the magnetic field. This decrease is accompanied  with the ground-state phase transition from $\vert 0,1,1 \rangle$ to $\vert 1,1,1\rangle_\text{I}$.
For the magnetic field larger that $|J_1|$, firstly, the negativity $N_\mathrm{AB}$  arises from zero and then continues to increase upon the increase of  $K$ until reaches fixed value $N_\mathrm{AB} \approx 0.207$. Under this conditions  the ground-state phase transition from $\vert 2,1,1 \rangle$ to $\vert 1,1,1\rangle_\text{I}$ happens. 
Here, any change in the behavior of bipartite entanglement negativity reminisces us the position of triple point at which three ground states $\vert 0,1,1 \rangle$, $\vert 1,1,1\rangle_\text{I}$ and $\vert 2,1,1 \rangle$ coexist together (see Fig. \ref{GSPT} (a)).
In result, we understand that  the cyclic four-spin exchange (even for $K\ll |J_1|$) plays an important role in the bipartite entanglement degree of the $\mathrm{Cu}_{4}^{\mathrm{II}}$ compound, too.

\subsubsection{Quantum teleportation and quantum Fisher information}\label{QFI}

Quantum teleportation is interpreted as transfer of quantum states from one place to another. The teleportating of a quantum bit (qubit) is achieved using quantum entanglement, in which two or more particles are indistinguishably connected to each other. If an entangled pair of particles are located in two spatially separate places, no matter the distance between them, the encoded quantum information can be teleported through a channel. The teleportation is the cornerstone of quantum networks, data storage, precision sensing and computing for ushering in a new era of high-secure communication.
In this section, we study the quantum teleportation throughout a couple of tetranuclear 
 $[\text{Cu}_4\text{L}_4(\text{H}_2\text{O})_4](\text{ClO}_4)_4$ complexes as an optimal resource of a generalized depolarizing channels \cite{Bowen2001}. In fact, the main goal of this section is to investigate the effects of 
cyclic four-spin exchange and the magnetic field on the possibility of teleportation through a couple of independent
 tetrameric Cu$^\text{II}_4$ systems. 
Let us assume the input state being an arbitrary unknown two-qubit pure state $\left\vert \psi _\text{in}\right\rangle =\cos (\frac{\theta }{2})\left\vert 10\right\rangle +e^{i\phi }\sin (\frac{\theta }{2})\left\vert 01\right\rangle$,
where $0\leq \theta \leq \pi $ and $0\leq \phi \leq 2\pi$. Angle $\theta$
describes an arbitrary state with specific amplitude and $\phi $ is the phase corresponds to this state.
The concurrence of the input state $\rho_\text{in}=\left\vert \psi _\text{in}\right\rangle \left\langle \psi _\text{in}\right\vert$
can be written as
\begin{equation}
{C}_\text{in}=2|e^{i\phi }\sin (\frac{\theta}{2})\cos (\frac{\theta }{2})|=|\sin (\theta)|.
\end{equation}%
Owing to the fact that quantum channels are known as 
completely positive and trace-preserving  (CPTP) operators,  
an input density operator is mapped to an output one \cite{Bowen2001}.
\begin{figure}
\centering
\includegraphics[scale=0.75,trim=110 120 10 90, clip]{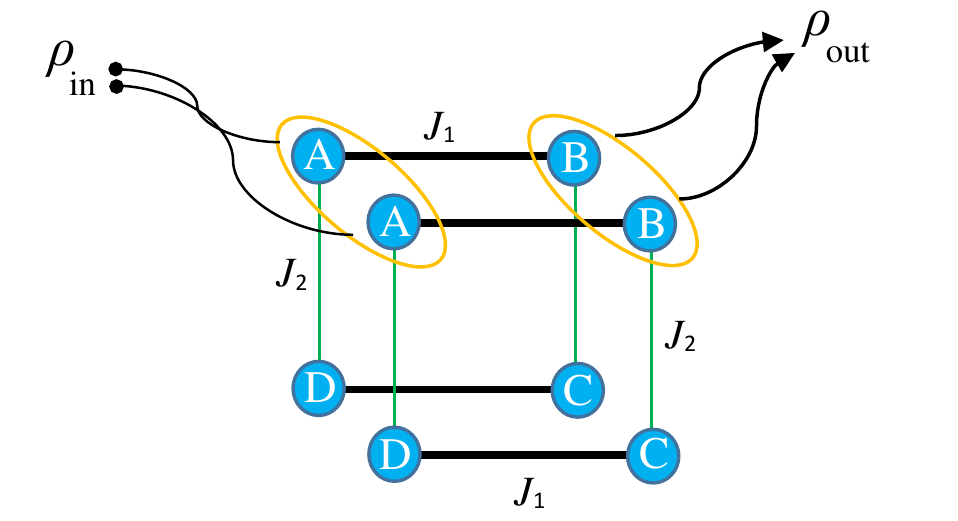} 
\caption{The schematic representation for the teleportation process of the input state $\rho_\mathrm{in}$ through a couple of discrete  tetrameric Cu$^\text{II}_4$ unit cells as quantum channels with density operator $\rho_\mathrm{ch}$. 
The teleported output state is demonstrated by ${\rho}_\mathrm{out}$.}
\label{QTele}
\end{figure}
Generally, when the quantum state is teleported through the mixed channel ${\rho }_\text{ch}$, the output replica state ${\rho }_\text{out}$ can be obtained by applying joint measurements and the local unitary
transformations on the input state $\rho _\text{in}$, hence
\begin{equation}
{\rho }_\text{out}=\sum_{i,j\in \{0,x,y,z\}}p_{i}p_{j}(\sigma ^{i}\otimes
\sigma ^{j})\rho _\text{in}(\sigma ^{i}\otimes \sigma ^{j}),
\end{equation}%
where $\sigma _{0}=I$  with $I$ being the $2\times 2$ identity matrix.
$p_{i}=\mathrm{Tr}[E^{i}\mathcal{\rho }_\text{ch}]$ satisfies the
condition $\sum_{i}p_{i}=1$. $E^{0}=\left\vert \Psi ^{-}\right\rangle
\left\langle \Psi ^{-}\right\vert ,$ $E^{1}=\left\vert \Phi
^{-}\right\rangle \left\langle \Phi ^{-}\right\vert ,$ $E^{2}=\left\vert
\Psi ^{+}\right\rangle \left\langle \Psi ^{+}\right\vert ,$ and $%
E^{3}=\left\vert \Phi ^{+}\right\rangle \left\langle \Phi ^{+}\right\vert ,$
from which $\{\left\vert \Psi ^{\pm }\right\rangle ,$ $\left\vert \Phi ^{\pm
}\right\rangle \}$ are Bell states.
 In this paper, we consider the density operator channel as ${\rho }_\text{ch}=\rho_\mathrm{AB}(T)$ 
 (see Appendix \ref{sec.appendixA}). 
 Therefore, the output density operator ${\rho }_\text{out}$
takes the form%
\begin{equation}
{\rho }_\text{out}=%
\begin{pmatrix}
\widetilde{{\rho }}_{11} & 0 & 0 & 0 \\
0 & \widetilde{{\rho }}_{22} & \widetilde{{\rho }}_{23}
& 0 \\
0 & \widetilde{{\rho }}_{23} & \widetilde{{\rho }}_{33}
& 0 \\
0 & 0 & 0 & \widetilde{{\rho }}_{11}%
\end{pmatrix}%
,
\end{equation}%
where%
\begin{eqnarray*}
\widetilde{{\rho }}_{11} &=&({\varrho}_{11}+{\varrho }_{44})({\varrho}_{22}+{\varrho}_{33}), \\
\widetilde{{\rho}}_{22} &=&({\varrho}_{11}+\varrho_{44})^{2}\cos^{2}(\frac{\theta}{2})+({\varrho}_{22}+
{\varrho}_{33})^{2}\sin^{2}(\frac{\theta}{2}), \\
\widetilde{{\rho }}_{33} &=&({\varrho }_{11}+{\varrho }_{44})^{2}\sin ^{2}(\frac{\theta }{2})+
({\varrho }_{22}+{\varrho }_{33})^{2}\cos ^{2}(\frac{\theta }{2}), \\
\widetilde{{\rho}}_{23} &=& 2e^{i\phi }\vert{\varrho }_{23}\vert^{2}\sin \theta.
\end{eqnarray*}

\begin{figure*}[tbp]
\centering
\includegraphics[width=8.25cm, height=6cm]{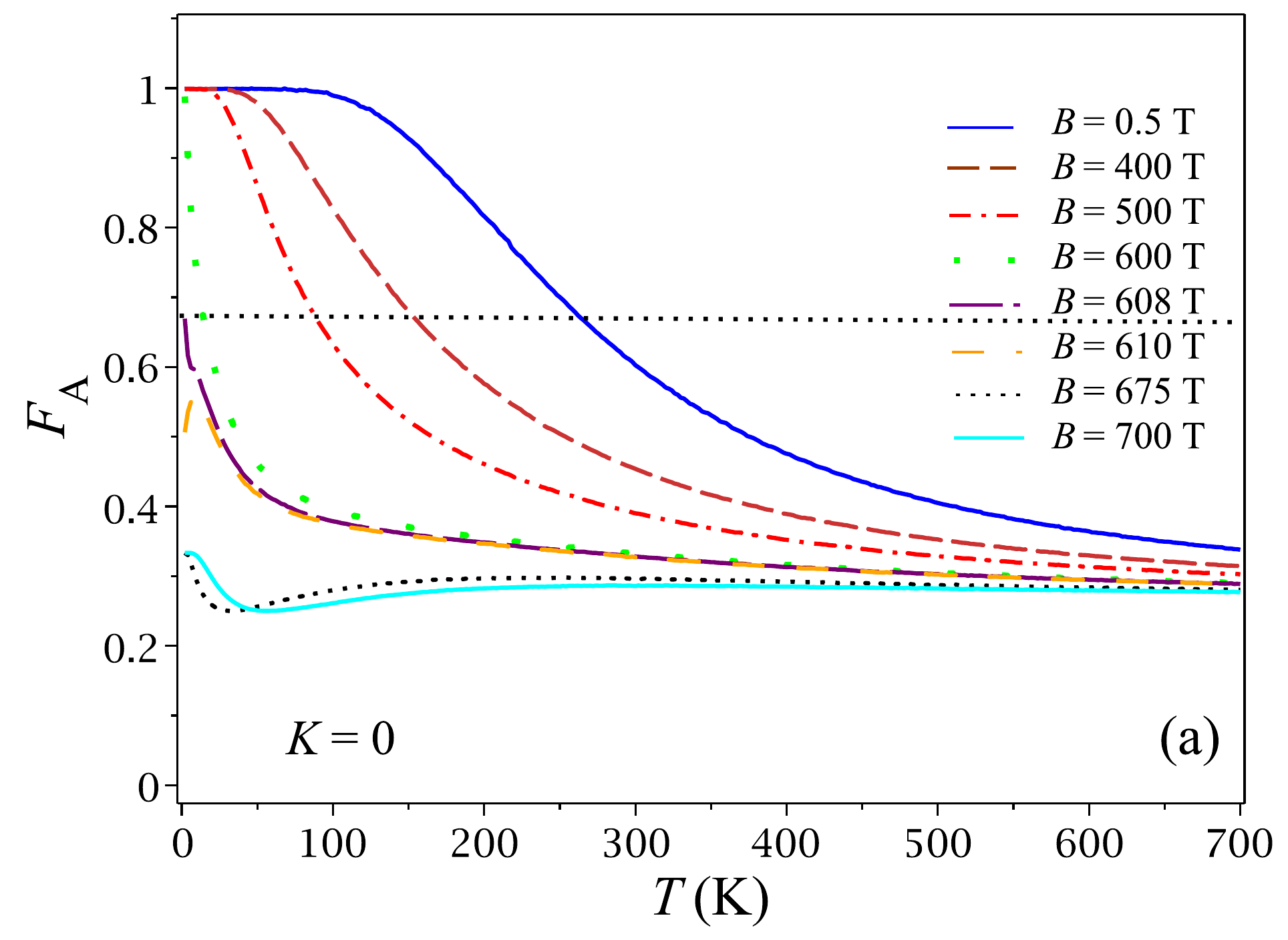} %
\includegraphics[width=8.25cm, height=6cm]{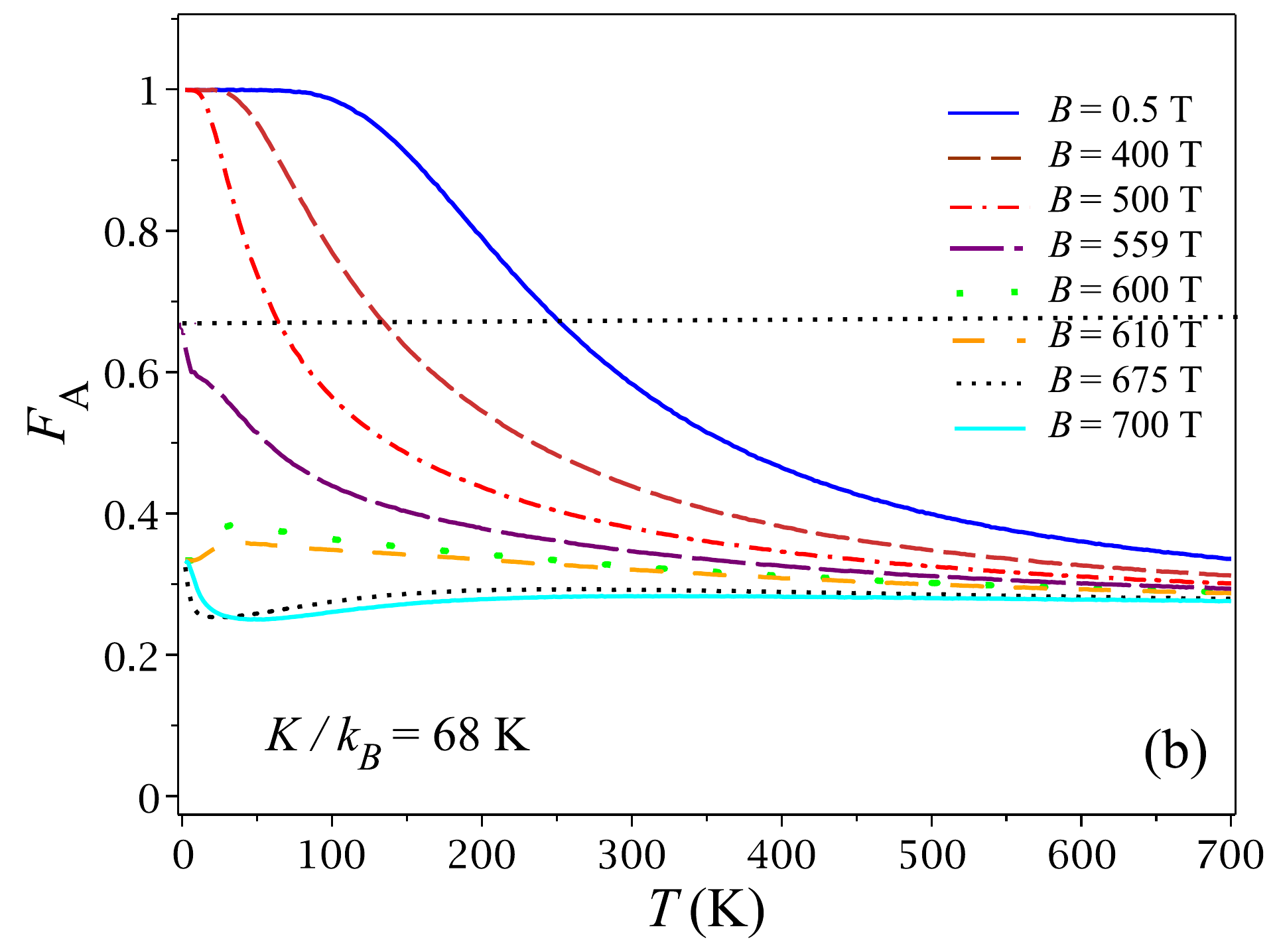}
\caption{Average fidelity ${F}_\text{A}$ versus temperature for parameter set $J_1/k_\text{B}=-638\; \text{K}$, $J_2/k_\text{B}=-34 \; \text{K}$ and fixed (a) $K=0$, (b) $K/k_\text{B}=68\; \text{K}$,  where several values of the magnetic field are considered.
}
\label{QTele}
\end{figure*}



To describe the quality of teleportation process, it is useful to study the fidelity between 
$\rho _\text{in}$ and ${\rho }_\text{out}$ to characterize the state that is teleported through a channel. When the input state is a pure state, one can apply the concept of fidelity as a tool to prob the teleportation performance of a quantum channel \cite{Jozsa1994}. The fidelity is defined by
\begin{equation}
{F}=\left(\mathrm{Tr}\sqrt{\sqrt{\rho _\text{in}}\mathcal{\rho}_\text{out}\sqrt{\rho _\text{in}}}\right)^{2},
\end{equation}%
thus in our case $F$ is deduced from the straightforward calculation
\begin{equation}
{F}=\frac{\sin ^{2}\theta }{2}[({\varrho }_{11}+{\varrho}_{44})^{2}+4{\varrho}_{23}^{2}-4{\varrho}_{22}^{2}]+4{\varrho}_{22}^{2}.
\end{equation}%
The average fidelity $\mathcal{F}_{A}$ can be calculated from the formula
\begin{equation}
{F}_\mathrm{A}=\frac{1}{4\pi }\int_{0}^{2\pi }d\phi
\int_{0}^{\pi }d\theta {F}\sin \theta.
  \label{fidelity}
\end{equation}%
After integrating Eq. (\ref{fidelity}), the average fidelity ${F}_\text{A}$
for the favorite bipartite subsystem can be given by
\begin{equation}
{F}_\text{A}=\frac{1}{3}[({\varrho }_{11}+{\varrho }_{44})^{2}+4{\varrho }_{23}^{2}-({\varrho}_{22}+
{\varrho }_{33})^{2}]+({\varrho }_{22}+{\varrho }_{33})^{2}.
\end{equation}%

By using quantum systems, scientists are seriously looking for a quantum state transmitting better than the classical communication protocols.
This superiority is achieved if and only if ${F}_\text{A}>2/3$ (horizontal dotted lines plotted in Fig. \ref{QTele}),
which ${F}_\text{A}=2/3$ is the best fidelity among the classical channels.
Let us proceed to examine the ability of system under study in teleporting information when the experimental data are considered.
In Fig. \ref{QTele} (a) the average fidelity $F_\mathrm{A}$ is shown as a function of the temperature for various magnetic fields such that exchange couplings are being $J_1/k_B=-638\; \text{K}$, $J_2/k_B=-34\; \text{K}$ when $K=0$.
 It can be comprehended from this figure that quantity ${F}_\mathrm{A}$ does not reaches the limit of quantum fidelities for
  $B\gtrsim 608 \; \text{T}$ that is comparable with the absolute value $|J_1|$, hence the teleportation of information happens for the magnetic field range $B<608\; \text{T}$. 
By looking at Fig. \ref{GSPT} (b), one understands that the quantum teleportation is solely possible for the area below than blue line at which a ground-state phase transition occurs between $\vert 1,1,1\rangle_\text{I}$ and  $\vert 0,1,1 \rangle$ states. 
The possibility of the teleportation decreases upon increase of the temperature.

In panel Fig. \ref{QTele} (b) we display the average fidelity versus the temperature for several magnetic fields and the same coupling constants values $J_1/k_B=-638\; \text{K}$, $J_2/k_B=-34\; \text{K}$ when the cyclic four-spin exchange term has 
nonzero value $K/k_B=68\; \text{K}$ (at which $\Pi_4=0$).
 Of course, for the magnetic fields remarkably lower than absolute value $|J_1|$ the fidelity is in its maximum value 
 ${F}_\mathrm{A}=1$ which monotonically decreases upon heating and tends to the constant value
 ${F}_\mathrm{A}=1/3$.
Considering the nonzero value of $K$ results in decreasing the ability of teleportation for magnetic fields close to $|J_1|$. In fact the critical magnetic field which determines the frontier between classical transmitting and quantum teleportation of information through this channel moves towards lower magnetic fields (see long-dashed purple lines of Figs. \ref{QTele} (a) and (b)).

 \begin{figure*}[tbp]
\centering
\includegraphics[width=8.75cm, height=6cm]{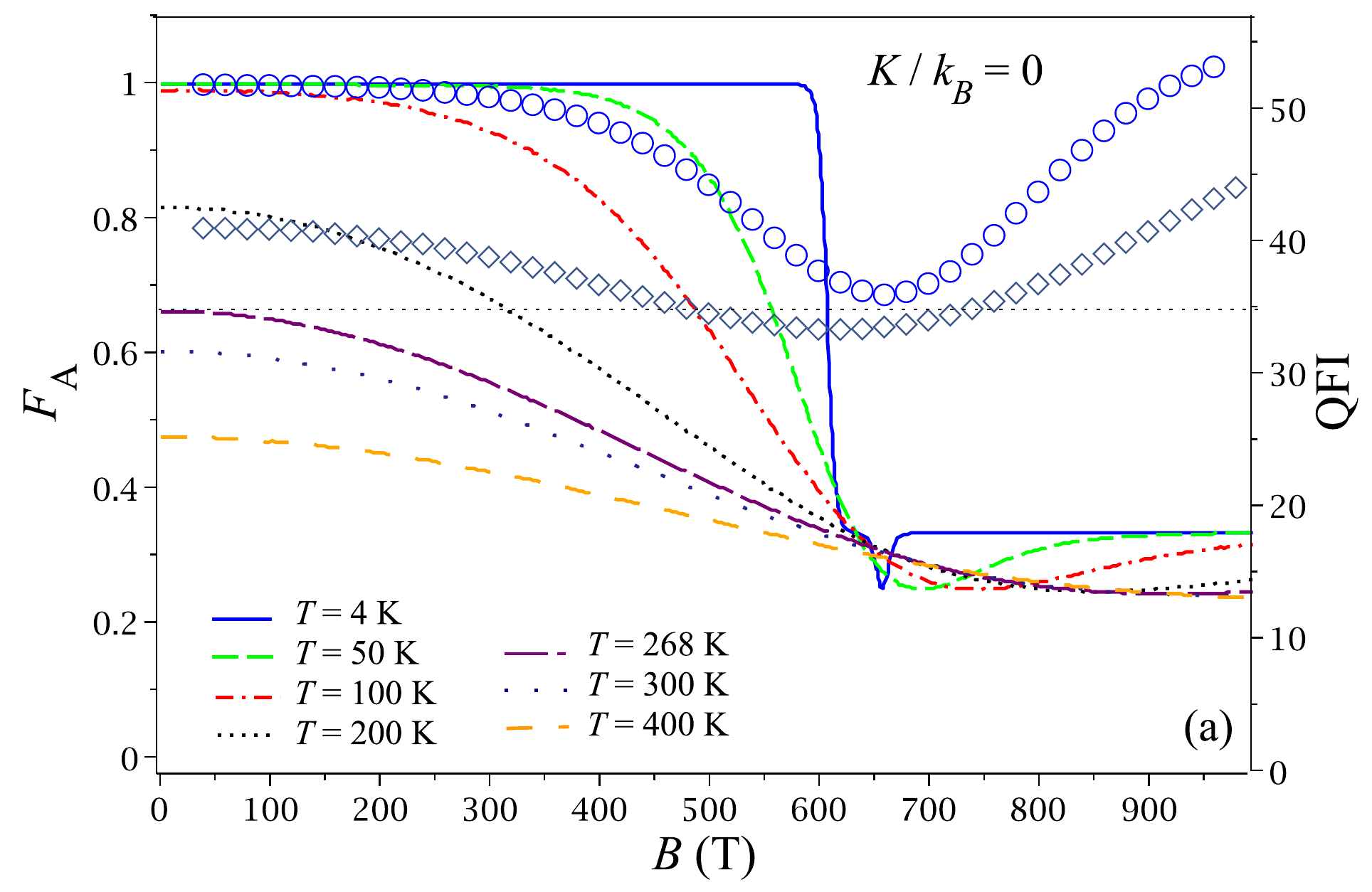} %
\includegraphics[width=8.75cm, height=6cm]{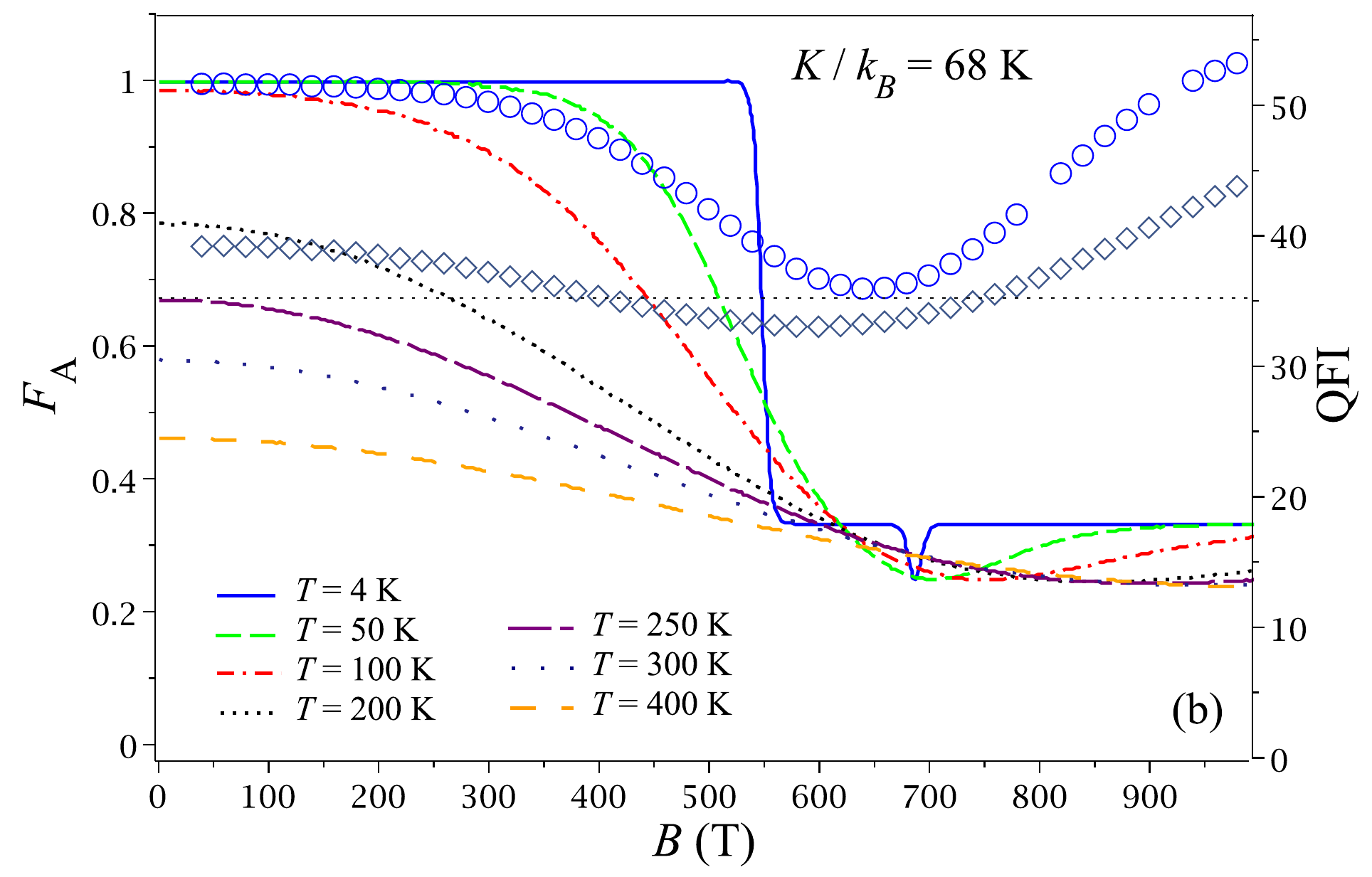}
\caption{
(a) Field dependence of the average fidelity ${F}_\text{A}$ for various temperatures, where other parameters have been taken as
  Fig. \ref{QTele} (a).
(b) The same function at  $K/k_\text{B}=68\; \text{K}$ against the magnetic field when different fixed values of the temperature are assumed. 
Horizontal dotted lines denote the limit value 2/3 and identify the quantum teleportation becomes possible for the case when the average fidelity is greater than this value. Circles and diamonds illustrated in panels (a) and (b) represent the QFI versus the magnetic field at  high temperature $T=100\; \text{K}$ and  $T=200\; \text{K}$ for the same set of other parameters to the corresponding $F_\text{A}$.}
\label{QTeleB}
\end{figure*}

To uncover the influence of the cyclic four-spin interaction on the magnetic field dependencies of average fidelity, a few plots
of $F_\text{A}$ versus the magnetic field are displayed in Figs. \ref{QTeleB} (a) and \ref{QTeleB} (b) for several temperatures by assuming  $K=0$ and $K/k_B=68\; \text{K}$, respectively.
 The first important thing that can be viewed in these figures is the existence of a sharp reentrance point in the average fidelity curve.
This phenomenon reminisces us the ground-state phase transition from $\vert 2,1,1 \rangle$ to $\vert 1,1,1\rangle_\text{I}$.
At low temperatures this quantity remains maximum (${F}_\mathrm{A}=1$) for notably high magnetic fields. 
Conventionally when the temperature increases the average fidelity decreases and the possibility of teleportation through this model is restricted.
The interesting point to emphasize is that, for the range $B<68\; \text{T}$ the average fidelity proves that the system finds the possibility of quantum teleportation close to the room temperature (see long-dashed purple line in Fig. \ref{QTeleB} (a)).
With increase of the cyclic four-spin interaction $K$ (see Fig. \ref{QTeleB} (b)) the average fidelity decreases and tends to classical limit 2/3. This reduction is more perspicuous close to the room temperature.

 Finally, we examine the effect of ring exchange $K$ on the QFI.
 For abbreviating the context, we have explained explicitly the QFI and its corresponding formulae for a bipartite spin-1/2 system in Appendix \ref{appendixB}.
In Figs. \ref{QTeleB} (a) and \ref{QTeleB} (b) the QFI alterations  are demonstrated versus the magnetic field for, respectively, $K=0$ and $K/k_B=68\; \text{K}$ at high temperatures $T=100\; \text{K}$ (circles) and $T=200\; \text{K}$ (diamonds). Exchange interactions have been considered antiferromagnetic as $J_1/k_B=-638\; \text{K}$ and $J_2/k_B=-34\; \text{K}$.
 This quantity gradually decreases from its maximum value with increase of the magnetic field and reaches a global minimum nearby the critical field that the average fidelity sharply decreases, then increases with further increase of the magnetic field. This behaviour guarantees the ground-state phase transition from $\vert 2,1,1\rangle$ to $\vert 1,1,1\rangle_\text{I}$.
The change of term $K$ leads to change the magnetic position of the minimum appeared in QFI curve.

\section{Conclusions}\label{conclusions}
 
 In  this work, we have considered a cyclic four-spin interaction for a
  spin-1/2 tetranuclear square-shaped complex whose nearest-neighbor sites interact with each other through the antiferromagnetic Heisenberg XXX exchange couplings $J_1$ and $J_2$. An external magnetic field in the $z-$direction has been hypothesized, as well. Next,  we have investigated the ground-state phase diagram of the model. The model exhibited four phases in its ground-state phase diagram, such a way that several discontinuous phase transitions together with a special triple point at which three different phases coexist together were observed.

Next, we have theoretically investigated the effects of the cyclic four-spin interaction on the whole entanglement degree of the spin-1/2 tetranuclear square-shaped model. We found that the cyclic four-spin interaction could notably enhance the whole entanglement negativity of the model. 
Afterwards, we have paid attention to the special case  tetranuclear square complex 
$[\text{Cu}_4\text{L}_4(\text{H}_2\text{O})_4](\text{ClO}_4)_4$ as an experimental representative of the spin-1/2 antiferromagnetic Heisenberg XXX model on a tetranuclear square compound \cite{Giri2011}. This complex is symmetric and possesses two different antiferromagnetic interactions among the copper ions such that one of them is much more stronger than another one. 
Then, we have theoretically examined for this system  the whole entanglement intensity and bipartite entanglement between pairs of Cu$^{\mathrm{II}}$ ions with stronger exchange interaction.
We have observed that considering a very small cyclic four-spin interaction may change the size of whole entanglement in this complex at moderate temperatures.
We have also realized that for the magnetic fields lower than the absolute value of the stronger coupling constant,  the whole entanglement of the system shows an unconventional minimum and a reentrant behavior at higher values of the cyclic four-spin exchange where a ground-state phase transition occurs. Increase of the cyclic four-spin term further than 
minimum point remarkably enhances the tetrapartite entanglement degree, where it does not vanish even at sufficiently high temperature.

Bipartite entanglement between pairs of Cu$^{\mathrm{II}}$ ions interacted through stronger exchange interaction remains in its maximum value for high temperatures and high magnetic fields.
 We understood that the bipartite entanglement remains robust at sufficiently high temperatures as large as the room temperature scale. At high temperature, bipartite entanglement is fragile against a cyclic four-spin interaction at high magnetic field. Indeed, increasing the cyclic four-spin exchange leads to decrease of the bipartite entanglement degree.

In addition, we have examined the possibility of teleportation through a couple of discrete tetranuclear copper $\text{Cu}_4^{\text{II}}$ complexes. We discovered that there is a critical magnetic field ($B\lesssim |J_1|$) at which the average fidelity goes further than classical limit 2/3 and the system gets the power of quantum teleportation of information. At weak magnetic field the model preserves its ability of teleportation at high enough temperatures.
 With increase of the cyclic four-spin term the average fidelity decays and the quantum teleportation is achievable at lower magnetic fields.
 
Finally, the QFI has been studied for this system by considering the actual values of the two-body exchange couplings.
  We have demonstrated that the QFI decreases from its maximum value upon increasing the magnetic field and reaches its minimum value close to the critical conditions at which the ground-state phase transition occurs between separable state and an entangled state.
The increase of the cyclic four-spin interaction results in decaying the QFI, as well.

\section*{Acknowledgments}
H. Arian Zad acknowledges the receipt of the grant from the Abdus Salam International Centre for Theoretical Physics (ICTP), Trieste, Italy. N. Ananikian  has been supported by the grant from CSMES RA in the frame of the research project No. SCS 21AG-1C0068.
H. Arian Zad and A. Zoshki acknowledge the receipt of the grant from the National scholarship Programme of the Slovak Republic (SAIA). M. Ja{\v s}{\v c}ur has been supported by the grant APVV-16-0186.
The authors are also grateful to Prof. Jozef Strečka for his insightful discussions.

\appendix

\section{Density matrices}\label{sec.appendixA}

The partition function of the tetranuclear $\mathrm{Cu}_{4}^{\mathrm{II}}$ square complex on the spin-1/2 XXX Heisenberg model containing two  coupling constants $J_1$ and $J_2$ and a cyclic four-spin exchange $K$ in the presence of an external magnetic field $B$ can be written as
\[
\begin{array}{lcl}
Z=2e^{-\frac{1}{4}\beta(\frac{1}{4}K-2(J_1+J_2))}\cosh{(2\beta\mu_BgB)}+
\\[0.1cm]
2\big[2e^{-\frac{1}{16}\beta K}\cosh(\frac{1}{2}\beta(J_1-J_2))+
e^{-\frac{1}{4}\beta(-\frac{7}{4}K+2(J_1+J_2))}+
\\[0.1cm]
e^{-\frac{1}{4}\beta(\frac{1}{4}K-2(J_1+J_2))}\big]\cosh{(\beta\mu_BgB)}+\\[0.1cm]
2e^{-\frac{1}{4}\beta(\frac{5}{4}K+2(J_1+J_2))}\cosh{(\beta\Lambda)}+
\\[0.1cm]
2e^{-\frac{1}{4}\beta(\frac{1}{4}K-2{J_2})}\cosh{(\frac{1}{2}\beta{J_1})}+
e^{-\beta(\frac{1}{16}K-\frac{1}{2}J_1+\frac{1}{2}J_2)}+\\[0.1cm]
e^{-\beta(-\frac{7}{16}K+\frac{1}{2}J_1+\frac{1}{2}J_2)}. \\[0.1cm]
\end{array}
\]
Function $\Lambda$ has already been introduced in the Sec. \ref{model}.
The nonzero coefficients of the tetramer density matrix $\rho_\mathrm{ABCD}$ are 
\[
\begin{array}{lcl}
\rho_{1,1},\rho_{2,2},\rho_{3,3},\rho_{4,4},\rho_{5,5},\rho_{6,6},\rho_{7,7},\rho_{8,8},\rho_{9,9},\rho_{10,10},\\
\rho_{11,11},\rho_{12,12},\rho_{13,13},\rho_{14,14},\rho_{15,15},\rho_{16,16},\\
\rho_{2,3},\rho_{2,5},\rho_{2,9},\rho_{3,5},\rho_{3,9},\rho_{4,6},\rho_{4,7},\rho_{4,10},\rho_{4,11},\rho_{4,13},\\
\rho_{5,9},\rho_{6,7},\rho_{6,10},\rho_{6,11},\rho_{6,13},\rho_{7,10},\rho_{7,11}, \rho_{7,13},\rho_{8,12},\\
\rho_{8,14},\rho_{8,15},\rho_{10,11},\rho_{10,13},\rho_{11,13},\rho_{12,14},\rho_{12,15},\rho_{14,15}
\end{array}
\]
where,  $\rho_{i,j}$ are long polynomials of parameters $J_1$, $J_2$, $K$ and $B$.
The nonzero coefficients of the density matrix $\varrho^\mathrm{A}=\rho^{T_\mathrm{A}}_\mathrm{ABCD}$ are 
\[
\begin{array}{lcl}
\varrho^\mathrm{A}_{i,i}=\rho_{i,i}\quad i=1\cdots 16,
\\[0.1cm]
\varrho^\mathrm{A}_{2,3}=\rho_{2,3},\varrho^\mathrm{A}_{2,5}=\rho_{2,5},\varrho^\mathrm{A}_{1,10}=\rho_{2,9},
\varrho^\mathrm{A}_{3,5}=\rho_{3,5},
\\[0.1cm]
\varrho^\mathrm{A}_{1,11}=\rho_{3,9},\varrho^\mathrm{A}_{4,6}=\rho_{4,6},\varrho^\mathrm{A}_{4,7}=\rho_{4,7},
\varrho^\mathrm{A}_{2,12}=\rho_{4,10},\\[0.1cm]
\varrho^\mathrm{A}_{3,12}=\rho_{4,11},\varrho^\mathrm{A}_{5,12}=\rho_{4,13},\varrho^\mathrm{A}_{1,13}=\rho_{5,9},
\varrho^\mathrm{A}_{6,7}=\rho_{6,7},
\\[0.1cm]
\varrho^\mathrm{A}_{2,14}=\rho_{6,10},\varrho^\mathrm{A}_{3,14}=\rho_{6,11},\varrho^\mathrm{A}_{5,14}=\rho_{6,13},
\varrho^\mathrm{A}_{2,15}=\rho_{7,10},
\\[0.1cm]
\varrho^\mathrm{A}_{3,15}=\rho_{7,11},\varrho^\mathrm{A}_{5,15}=\rho_{7,13},\varrho^\mathrm{A}_{4,16}=\rho_{8,12},
\varrho^\mathrm{A}_{6,16}=\rho_{8,14},
\\[0.1cm]
\varrho^\mathrm{A}_{7,16}=\rho_{8,15},\varrho^\mathrm{A}_{10,11}=\rho_{10,11},\varrho^\mathrm{A}_{10,13}=\rho_{10,13},
\\[0.1cm]
\varrho^\mathrm{A}_{11,13}=\rho_{11,13},\varrho^\mathrm{A}_{12,14}=\rho_{12,14},\varrho^\mathrm{A}_{12,15}=\rho_{12,15},
\\[0.1cm]
\varrho^\mathrm{A}_{14,15}=\rho_{14,15}
\end{array}
\]

The nonzero coefficients of the density matrix $\varrho^\mathrm{B}=\rho^{T_\mathrm{B}}_\mathrm{ABCD}$ are 
\[
\begin{array}{lcl}
\varrho^\mathrm{B}_{i,i}=\rho_{i,i}\quad i=1\cdots 16,
\\[0.1cm]
\varrho^\mathrm{B}_{2,3}=\rho_{2,3},\varrho^\mathrm{B}_{1,6}=\rho_{2,5},\varrho^\mathrm{B}_{2,9}=\rho_{2,9},\varrho^\mathrm{B}_{1,7}=\rho_{3,5},\varrho^\mathrm{B}_{3,9}=\rho_{3,9},
\\[0.1cm]
\varrho^\mathrm{B}_{4,6}=\rho_{4,6},\varrho^\mathrm{B}_{4,7}=\rho_{4,7},\varrho^\mathrm{B}_{4,10}=\rho_{4,10},
\varrho^\mathrm{B}_{4,11}=\rho_{4,11},
\\[0.1cm]
\varrho^\mathrm{B}_{4,13}=\rho_{4,13},\varrho^\mathrm{B}_{1,13}=\rho_{5,9},
\varrho^\mathrm{B}_{6,7}=\rho_{6,7},\varrho^\mathrm{B}_{6,10}=\rho_{6,10},
\\[0.1cm]
\varrho^\mathrm{B}_{6,11}=\rho_{6,11},\varrho^\mathrm{B}_{6,13}=\rho_{6,13},\varrho^\mathrm{B}_{7,10}=\rho_{7,10},
\varrho^\mathrm{B}_{7,11}=\rho_{7,11},
\\[0.1cm]
\varrho^\mathrm{B}_{7,13}=\rho_{7,13},\varrho^\mathrm{B}_{8,12}=\rho_{8,12},\varrho^\mathrm{B}_{8,14}=\rho_{8,14},\varrho^\mathrm{B}_{8,15}=\rho_{8,15},
\\[0.1cm]
\varrho^\mathrm{B}_{10,11}=\rho_{10,11},\varrho^\mathrm{B}_{9,14}=\rho_{10,13},
\varrho^\mathrm{B}_{9,15}=\rho_{11,13},\varrho^\mathrm{B}_{10,16}=\rho_{12,14},
\\[0.1cm]
\varrho^\mathrm{B}_{11,16}=\rho_{12,15},
\varrho^\mathrm{B}_{14,15}=\rho_{14,15}
\end{array}
\]

The nonzero coefficients of the density matrix $\varrho^\mathrm{C}=\rho^{T_\mathrm{C}}_\mathrm{ABCD}$ are 
\[
\begin{array}{lcl}
\varrho^\mathrm{C}_{i,i}=\rho_{i,i}\quad i=1\cdots 16,\\[0.1cm]
\varrho^\mathrm{C}_{1,4}=\rho_{2,3},\varrho^\mathrm{C}_{2,5}=\rho_{2,5},\varrho^\mathrm{C}_{2,9}=\rho_{2,9},
\varrho^\mathrm{C}_{1,7}=\rho_{3,5},\varrho^\mathrm{C}_{1,11}=\rho_{3,9},
\\[0.1cm]
\varrho^\mathrm{C}_{2,8}=\rho_{4,6},\varrho^\mathrm{C}_{4,7}=\rho_{4,7},\varrho^\mathrm{C}_{2,12}=\rho_{4,10},
\varrho^\mathrm{C}_{4,11}=\rho_{4,11},
\\[0.1cm]
\varrho^\mathrm{C}_{2,15}=\rho_{4,13},\varrho^\mathrm{C}_{5,9}=\rho_{5,9},\varrho^\mathrm{C}_{5,8}=\rho_{6,7},
\varrho^\mathrm{C}_{6,10}=\rho_{6,10},
\\[0.1cm]
\varrho^\mathrm{C}_{8,9}=\rho_{6,11},\varrho^\mathrm{C}_{6,13}=\rho_{6,13},\varrho^\mathrm{C}_{5,12}=\rho_{7,10},
\varrho^\mathrm{C}_{7,11}=\rho_{7,11},
\\[0.1cm]
\varrho^\mathrm{C}_{5,15}=\rho_{7,13},\varrho^\mathrm{C}_{8,12}=\rho_{8,12},
\varrho^\mathrm{C}_{6,16}=\rho_{8,14},
\\[0.1cm]
\varrho^\mathrm{C}_{8,15}=\rho_{8,15},\varrho^\mathrm{C}_{9,12}=\rho_{10,11},\varrho^\mathrm{C}_{10,13}=\rho_{10,13},
\varrho^\mathrm{C}_{9,15}=\rho_{11,13},
\\[0.1cm]
\varrho^\mathrm{C}_{10,16}=\rho_{12,14},\varrho^\mathrm{C}_{12,15}=\rho_{12,15},\varrho^\mathrm{C}_{13,16}=\rho_{14,15}
\end{array}
\]

The nonzero coefficients of the density matrix $\varrho^\mathrm{D}=\rho^{T_\mathrm{D}}_\mathrm{ABCD}$ are 
\[
\begin{array}{lcl}
\varrho^\mathrm{D}_{i,i}=\rho_{i,i}\quad i=1\cdots 16,
\\[0.1cm]
\varrho^\mathrm{D}_{1,4}=\rho_{2,3},\varrho^\mathrm{D}_{1,6}=\rho_{2,5},\varrho^\mathrm{D}_{1,10}=\rho_{2,9},
\varrho^\mathrm{D}_{3,5}=\rho_{3,5},\varrho^\mathrm{D}_{3,9}=\rho_{3,9},
\\[0.1cm]
\varrho^\mathrm{D}_{2,8}=\rho_{4,6},\varrho^\mathrm{D}_{3,8}=\rho_{4,7},\varrho^\mathrm{D}_{4,10}=\rho_{4,10},
\varrho^\mathrm{D}_{3,12}=\rho_{4,11},
\\[0.1cm]
\varrho^\mathrm{D}_{3,14}=\rho_{4,13},\varrho^\mathrm{D}_{5,9}=\rho_{5,9},\varrho^\mathrm{D}_{5,8}=\rho_{6,7},
\varrho^\mathrm{D}_{6,10}=\rho_{6,10},
\\[0.1cm]
\varrho^\mathrm{D}_{5,12}=\rho_{6,11},\varrho^\mathrm{D}_{5,14}=\rho_{6,13},\varrho^\mathrm{D}_{8,9}=\rho_{7,10},
\varrho^\mathrm{D}_{7,11}=\rho_{7,11},
\\[0.1cm]
\varrho^\mathrm{D}_{7,13}=\rho_{7,13},\varrho^\mathrm{D}_{8,12}=\rho_{8,12},\varrho^\mathrm{D}_{8,14}=\rho_{8,14},
\varrho^\mathrm{D}_{7,16}=\rho_{8,15},
\\[0.1cm]
\varrho^\mathrm{D}_{9,12}=\rho_{10,11},\varrho^\mathrm{D}_{9,14}=\rho_{10,13},
\varrho^\mathrm{D}_{11,13}=\rho_{11,13},\varrho^\mathrm{D}_{12,14}=\rho_{12,14},
\\[0.1cm]
\varrho^\mathrm{D}_{11,16}=\rho_{12,15},
\varrho^\mathrm{D}_{13,16}=\rho_{14,15},
\\[0.4cm]
\end{array}
\]

The explicit nonzero elements of the reduced density matrix $\rho_\mathrm{AB}$ (\ref{state2}) are 

\[
\begin{array}{lcl}
\varrho_{1,1}=\rho_{1,1}+\rho_{2,2}+\rho_{3,3}+\rho_{4,4},\\[0.1cm]
\varrho_{2,2}=\rho_{5,5}+\rho_{6,6}+\rho_{7,7}+\rho_{8,8},\\[0.1cm]
\varrho_{3,3}=\rho_{9,9}+\rho_{10,10}+\rho_{11,11}+\rho_{12,12},\\[0.1cm]
\varrho_{4,4}=\rho_{13,13}+\rho_{14,14}+\rho_{15,15}+\rho_{16,16},\\[0.1cm]
\varrho_{2,3}=\varrho_{2,3}^{*}=\rho_{5,9}+\rho_{6,10}+\rho_{7,11}+\rho_{8,12}.
\end{array}
\]

The nonzero coefficients of the reduced density matrix $\Omega^\mathrm{A}=\rho^{T_\mathrm{A}}_\mathrm{AB}$ are 

\[
\begin{array}{lcl}
\Omega^\mathrm{A}_{1,1}=\rho_{1,1}+\rho_{2,2}+\rho_{3,3}+\rho_{4,4},\\[0.1cm]
\Omega^\mathrm{A}_{2,2}=\rho_{5,5}+\rho_{6,6}+\rho_{7,7}+\rho_{8,8},\\[0.1cm]
\Omega^\mathrm{A}_{3,3}=\rho_{9,9}+\rho_{10,10}+\rho_{11,11}+\rho_{12,12},\\[0.1cm]
\Omega^\mathrm{A}_{4,4}=\rho_{13,13}+\rho_{14,14}+\rho_{15,15}+\rho_{16,16},\\[0.1cm]
\Omega^\mathrm{A}_{1,4}=\Omega^\mathrm{A}_{4,1}=\rho_{5,9}+\rho_{6,10}+\rho_{7,11}+\rho_{8,12}.
\end{array}
\]

The nonzero coefficients of the reduced density matrix $\Omega^\mathrm{A}=\rho^{T_\mathrm{A}}_\mathrm{AC}$ are

\[
\begin{array}{lcl}
\Omega^\mathrm{A}_{1,1}=\rho_{1,1}+\rho_{2,2}+\rho_{5,5}+\rho_{6,6},\\[0.1cm]
\Omega^\mathrm{A}_{2,2}=\rho_{3,3}+\rho_{4,4}+\rho_{7,7}+\rho_{8,8},\\[0.1cm]
\Omega^\mathrm{A}_{3,3}=\rho_{9,9}+\rho_{10,10}+\rho_{13,13}+\rho_{14,14},\\[0.1cm]
\Omega^\mathrm{A}_{4,4}=\rho_{11,11}+\rho_{12,12}+\rho_{15,15}+\rho_{16,16},\\[0.1cm]
\Omega^\mathrm{A}_{1,4}=\Omega^\mathrm{A}_{4,1}=\rho_{3,9}+\rho_{4,10}+\rho_{7,13}+\rho_{8,14}.
\end{array}
\]

The nonzero coefficients of the reduced density matrix $\Omega^\mathrm{B}=\rho^{T_\mathrm{B}}_\mathrm{BC}$ are

\[
\begin{array}{lcl}
\Omega^\mathrm{B}_{1,1}=\rho_{1,1}+\rho_{2,2}+\rho_{9,9}+\rho_{10,10},\\[0.1cm]
\Omega^\mathrm{B}_{2,2}=\rho_{3,3}+\rho_{4,4}+\rho_{11,11}+\rho_{12,12},\\[0.1cm]
\Omega^\mathrm{B}_{3,3}=\rho_{5,5}+\rho_{6,6}+\rho_{13,13}+\rho_{14,14},\\[0.1cm]
\Omega^\mathrm{B}_{4,4}=\rho_{7,7}+\rho_{8,8}+\rho_{15,15}+\rho_{16,16},\\[0.1cm]
\Omega^\mathrm{B}_{1,4}=\Omega^\mathrm{B}_{4,1}=\rho_{3,5}+\rho_{4,6}+\rho_{11,13}+\rho_{12,14}.\\[0.2cm]
\end{array}
\]

\section{QFI}\label{appendixB}
We here verify the QFI for the pair spins A and B.
The alteration of the thermal quantum state ${\rho}_\mathrm{AB}(T)$ (\ref{state2}) under a unitary transformation can be described as 
${\rho}(\Theta)=\exp[-i\mathcal{A}\Theta]{\rho}_\mathrm{AB}\exp[i\mathcal{A}\Theta]$, where $\Theta$ is the phase shift and
 $\mathcal{A}$ is an operator. The estimation accuracy for $\Theta$ is limited by the quantum
Cram{\' e}r-Rao inequality $\Delta\hat{\Theta}\geq \frac{1}{\sqrt{\nu\mathcal{F}({\rho}_{\Theta})}}$,
where $\hat{\Theta}$ is the unbiased estimator for $\Theta$, and $\nu$ is the
number of measurement repetitions. Therefore, the QFI of the given state can be expressed as \cite{Holevo,Liu2013}
\begin{equation} \label{eq:QFI}
\begin{array}{cl} 
 \mathcal{F}({\rho}_\mathrm{AB}, \mathcal{A})=2\sum\limits_{i,j=1}^4\frac{(\tau_i-\tau_j)^2}{\tau_i+\tau_j}
 \mid\langle \chi_i\mid\mathcal{A}\mid\chi_j\rangle\mid^2,
\end{array}
\end{equation}
where $\vert\chi_i\rangle$ and $\tau_i$ are, respectively, the eigenvectors and their corresponding eigenvalues of
 the density matrix ${\rho}_\mathrm{AB}$.
The complete sets of local orthonormal observables $\{\mathcal{A}_{\eta}\}$ and $\{\mathcal{B}_{\eta}\}$ corresponding to the  subsystems, respectively, $\mathrm{A}$ and $\mathrm{B}$ might be employed to quantify QFI of their states ${\rho}_\mathrm{AB}$. 
The QFI for a general bipartite spin-1/2 system reads \cite{Li2013}
\begin{equation} \label{eq:QFI_2}
\begin{array}{cl} 
 \mathcal{F}=\sum\limits_{\eta} \mathcal{F}(\tilde{\rho}, \mathcal{A}_{\eta}\otimes I + I\otimes \mathcal{B}_{\eta}).
\end{array}
\end{equation}
The local observables $\{\mathcal{A}_{\eta}\}$ and $\{\mathcal{B}_{\eta}\}$ are orthonormal and can be written as
\begin{equation} \label{eq:AB}
\begin{array}{cl} 
 \{\mathcal{A}_{\eta}\}=\{\mathcal{B}_{\eta}\}=\sqrt{2}\{I,\;S^x,\;S^y,\;S^z\},
\end{array}
\end{equation}
which $I$ is $2\times 2$ identity matrix. Ultimately, by substituting Eq. (\ref{eq:AB}) to Eq. (\ref{eq:QFI}), QFI $\mathcal{F}$ is  achieved.

\section*{}

\bigskip

\end{document}